\documentclass[prd,preprint,superscriptaddress]{revtex4}
\pdfoutput=1
\usepackage{revsymb}
\usepackage{amsmath}
\usepackage[pdftex]{graphicx}
\usepackage{graphicx}
\usepackage{bm}
\usepackage{subfigure}
\usepackage{slashbox}
\usepackage{epstopdf}
\DeclareGraphicsExtensions{.pdf,.png,.jpg}
\newcommand{\be}{\begin{equation}}
\newcommand{\ee}{\end{equation}}
\newcommand{\ben}{\begin{eqnarray}}
\newcommand{\een}{\end{eqnarray}}

\begin{document}
\title{Perturbations for transient acceleration}

\author{Cristofher Zu\~niga Vargas}
\email{win_unac@hotmail.com}
\affiliation{Universidade Federal do Esp\'{\i}rito Santo,
Departamento
de F\'{\i}sica\\
Av. Fernando Ferrari, 514, Campus de Goiabeiras, CEP 29075-910,
Vit\'oria, Esp\'{\i}rito Santo, Brazil}

\author{Wiliam S. Hip\'{o}lito-Ricaldi}
\email{hipolito@ceunes.ufes.br}
\affiliation{Universidade Federal do Esp\'{\i}rito Santo, Departamento de Ci\^encias Naturais,
Grupo de F\'{\i}sica Te\'{o}rica, \\
Rodovia BR 101 Norte, km 60, Campus de S\~{a}o Mateus,
 CEP 29932-540,
S\~ao Mateus, Esp\'{\i}rito Santo, Brazil}

\author{Winfried Zimdahl}
\email{winfried.zimdahl@pq.cnpq.br}
\affiliation{Universidade Federal do Esp\'{\i}rito Santo,
Departamento
de F\'{\i}sica\\
Av. Fernando Ferrari, 514, Campus de Goiabeiras, CEP 29075-910,
Vit\'oria, Esp\'{\i}rito Santo, Brazil}
\date{\today}
\date{\today}

\begin{abstract}
According to the standard $\Lambda$CDM model, the accelerated expansion of the Universe will go on forever. Motivated by recent observational results, we explore the possibility of a finite phase of acceleration which asymptotically approaches another period of decelerated expansion. Extending an earlier study on a corresponding homogeneous and isotropic dynamics, in which interactions between dark matter and dark energy are crucial, the present paper also
investigates the dynamics of the matter perturbations both on the Newtonian and General Relativistic (GR)  levels and
quantifies the potential relevance of perturbations of the dark-energy component.
In the background, the model is tested against the Supernova type Ia (SNIa) data of the Constitution set
and on the perturbative level against growth rate data, among them those of the WiggleZ survey, and the data of the 2dFGRS project.
Our results indicate that a transient phase of accelerated expansion is not excluded by current observations.
\end{abstract}
\maketitle

\section{Introduction}

It is generally believed that our presently observable Universe is dynamically dominated by a dark sector which is composed of a dark-energy component with a large negative pressure and pressureless dark matter. The physical nature of both these components remains a mystery, notwithstanding the intense research activity in the field since the discovery of the accelerated expansion of the Universe in
\cite{riessperl}. Direct and indirect support for this result has been accumulating over the past years. This comprises further results from advanced data sets for supernovas of type Ia (SNIa) as well as results from studies of the large scale
structure \cite{lss}, cosmic microwave background \cite{cmb}, the
integrated Sachs--Wolfe effect \cite{isw}, baryonic acoustic
oscillations \cite{eisenstein} and  gravitational lensing
\cite{weakl}.
By now, there exists a standard model, the $\Lambda$CDM model, which, \textit{grosso modo}, is compatible with the cosmological data. (Notice, however, that there is an ongoing discussions, see, e.g., \cite{nieuwenh}, on apparent shortcomings of this model).
Nevertheless, because of the cosmological constant problem in its different facets, including the coincidence problem, a still growing number of competing models has been developed over the last years, most of them ``dynamizing" the cosmological constant or even generalizing Einstein's theory.
Observations force these models to have  a dynamics that is very similar to that of the $\Lambda$CDM model around the present epoch. Moreover, the past evolution is restricted by the necessity of a matter-dominated epoch to guarantee cosmic structure formation. The future cosmological evolution within alternative models, however, may be very different from a de Sitter phase, which is the final fate of a
$\Lambda$CDM universe and also of several other approaches like Chaplygin-gas scenarios.
Phantom-type dark energy with a constant equation-of-state parameter, e.g., will end in a big-rip singularity after a finite time \cite{caldwell,caldwell2}. More recently, a still different scenario, called ``little rip", was proposed \cite{frampton}.
Already at the beginning of the past decade several authors discussed the possibility that the currently observed accelerated expansion might be a transient phenomenon, i.e. that there might occur a transition back to decelerated expansion \cite{albrecht,barrow,bertolami}.
Some recent observations seem to back up this idea.
Evidence was found for a slowing-down of the expansion rate of the Universe, equivalent to an increase of the deceleration parameter $q(z)$ for decreasing redshifts $z$ close to the present epoch $z=0$ 
\cite{sastaro,antonio,liwuyu,caituo}.
This could indicate a scenario, according to which the observed accelerated expansion of the Universe is a transient phenomenon, implying a transition back to a decelerated expansion either for the future evolution with $z\lesssim 0$ or even around the current epoch $z\gtrsim 0$.
Scenarios of transient acceleration were previously discussed in \cite{alcaniz} and  \cite{alcaniztr}.
The model on which the present paper relies was developed in \cite{transient}.
It describes transient cosmological acceleration as the consequence of an interaction between dark matter and dark energy. Such a dynamics cannot be obtained if the interaction represents a small correction to the standard $\Lambda$CDM model. For models of such type  the long-time cosmological dynamics will always be determined by the cosmological term and result
in accelerated expansion. To achieve transient accelerated expansion, a twofold role of the interaction is necessary. At first, it has to cancel the ``bare" cosmological constant and at second it has to generate a phase of accelerated expansion by itself. Acceleration has to be an interaction phenomenon.
As it was shown in \cite{transient}, these requirements can be fulfilled by interaction terms that combine powers and exponentials of the cosmic scale factor. Even though this specific choice was made for mathematical convenience, we expect that the mentioned two features will be crucial for a broader class of models.
While the study in \cite{transient} was restricted to the homogeneous and isotropic background,
the present paper investigates the corresponding perturbation dynamics as well. In particular, we calculate the growth rate of the matter perturbations and the matter power spectrum.
We compare our results with the growth-rate data collected in \cite{gong}
as well as with those of the  WiggleZ survey \cite{blake}
and with the data from the
2dFGRS program \cite{cole}.
Emphasis is also put on the potential relevance of perturbations of the dark-energy component
which are neglected by many studies of the matter perturbation behavior (for exceptions see, e.g. \cite{sapone,Park-Hwang,vernizzi}). In concordance with
parallel investigations for other models \cite{saulo}, we find that they are small indeed on scales that are relevant for structure formation.
However, there are indication that their role is increasing with increasing scale.
The background dynamics is reconsidered on the basis of the SNIa observations by the Constitution set \cite{Hicken}.

The paper is organized as follows. In section \ref{model}  we reanalyze  the basic features and the homogeneous and isotropic background dynamics of the transient acceleration model. Section \ref{newton} is devoted to a Newtonian treatment of the perturbation dynamics. A fully relativistic, gauge-invariant
investigation and a calculation of the matter power spectrum are the subjects of section \ref{rel}. In section \ref{summary} we summarize and discuss our results.

\section{The transient acceleration model}
\label{model}

We assume the cosmic substratum to be dynamically dominated by a mixture of a pressureless matter fluid and a dark-energy component.
The field equations for a spatially flat, homogeneous and isotropic
two-component universe of this type are the Friedmann  equation
\begin{equation}
3\,H^{2} = 8\,\pi\,G\,\left(\rho_{m} + \rho_{x}\right)\
\label{friedmann}
\end{equation}
and
\begin{equation}
\dot{H}\, = - 4\,\pi\,G\,\left(\rho_{m} + \rho_{x} + p_{x}\right)
\ .\label{dotH}
\end{equation}
Here, $\rho_{m}$ is the energy density of pressureless dark matter
and $\rho_{x}$ is
the density of the dark-energy component with a pressure $p_{x}$. The Hubble rate $H$ is given by
$H = \frac{\dot{a}}{a}$, where $a$ is the scale factor of the Robertson-Walker metric and a dot denotes the derivative with respect to the cosmic time.
We assume
that both the dark components do not conserve separately but interact with
each other in such a manner that the balance equations take the
form
\begin{equation}
\dot{\rho}_{m} + 3H \rho_{m} = Q \,  \label{dotrhom}
\end{equation}
and
\begin{equation}
\dot{\rho}_{x} + 3H (1+w)\rho_{x} = - Q \, , \label{dotrhox}
\end{equation}
where $w\equiv \frac{p_{x}}{\rho_{x}}$ is the equation-of-state parameter of the dark energy.
The sum of (\ref{dotrhom}) and (\ref{dotrhox}) results in the total energy conservation equation
$\dot{\rho} + 3H \left(\rho + p\right) = 0$,
where the total pressure
equals the dark energy pressure, $p = p_{X}$.

It is convenient to write the energy density of the matter fluid
as
\begin{equation}
\rho_m=\tilde\rho_{m_0}a^{-3}\,f\left(a\right)\ ,
\label{materia}
\end{equation}
where we have chosen $a_{0}=1$ for the present value of the scale factor.
The quantity $\tilde\rho_{m_0}$ is a constant and $f(a)$ is an
arbitrary time-dependent function. With
$f\left(a\right)=1+g\left(a\right)$, this structure implies that
\begin{equation}\label{defQ}
Q=\rho_m\frac{\dot{f}}{f}=\tilde\rho_{m_0}a^{-3}\dot{f} = \tilde\rho_{m_0}\frac{d g}{da}\dot{a}a^{-3} \
\end{equation}
and
\begin{equation}
\rho_m=\tilde\rho_{m_0}\left(1+g\right)a^{-3}\ .
\label{rm}
\end{equation}
The present values $\rho_{m_{0}}$ and $\tilde\rho_{m_{0}}$ are related by
\begin{equation}
{\rho}_{m_{0}} = \tilde\rho_{m_{0}}\left(1 + g_{0}\right) \ ,
\label{tilm0}
\end{equation}
where $g_{0} \equiv g(1)$.
The quantity
${\rho}_{m_{0}}$ is the value of $\rho_{m}$ at $a=1$ in
the presence of the interaction, $\tilde\rho_{m_{0}}$ is the value of
$\rho_{m}$ at $a=1$ for vanishing interaction. The interaction
re-normalizes the present value of $\rho_{m}$.

In \cite{transient} it has been shown that an analytically solvable transient acceleration scenario can be based
on an equation-of-state parameter $w=-1$ with an interaction, characterized by
\begin{equation}
g(a)=c a^5 \exp (-a^2/\sigma ^2) \ .
\label{g}
\end{equation}
In the following, we briefly recall the basic features of this approach.
We start by integrating equation (\ref{dotrhom}) with (\ref{defQ}) and (\ref{g}) which yields
\begin{equation}
\label{rhom2}
\rho_{m} = \rho_{m_0}a^{-3}  + Ka^{-3}\left[a^{5}\exp(-a^{2}/\sigma^{2}) - \exp(-1/\sigma^{2})\right]\ ,
\end{equation}
where $K\equiv c\tilde\rho_{m_0}$, while from (\ref{dotrhox}), (\ref{defQ}) and (\ref{g}) it follows that
\begin{equation}
\rho_{x} = \rho_{x_{0}}^{eff}
- K\,\exp\left(-a^{2}/\sigma^{2}\right) \left(a^{2} - \frac{3}{2}\sigma^{2}\right)
\ ,
\label{rh2}
\end{equation}
and
\begin{equation}
\rho_{x_{0}}^{eff}
= \rho_{x_{0}} - \frac{3}{2}K\,\exp (-1/\sigma^{2})
\left[\sigma^{2} - \frac{2}{3}\right]\ .
\label{reffh2}
\end{equation}
In the interaction-free limit $K\rightarrow 0$ we have consistently $\rho_{x}\rightarrow \rho_{x_{0}} =$ const. The quantity $\rho_{x_{0}}^{eff}$ can be seen as an effective cosmological constant which is re-normalized compared with the ``bare" value, corresponding to $\rho_{x_{0}}$, due to the presence of an interaction.
The ratio $\frac{\ddot{a}}{a}$ becomes
\begin{equation}
\label{accOmega2}
\frac{\ddot a}{a}=-\frac{1}{2}H_{0}^{2}
\left\{\frac{\Omega_{m_{0}} - \bar{K}\exp (-1/\sigma^{2})}{a^3}  -2\Omega_{x_0}^{eff} + 3 \bar{K} \exp
(-a^{2}/\sigma^{2}) \left[a^{2} - \sigma^{2}\right]\right\}
\end{equation}
with $\bar{K} = \frac{8 \pi G}{3 H_{0}^{2}}K$ and $\Omega_{x_0}^{eff} = \frac{8 \pi G}{3 H_{0}^{2}}\rho_{x_0}^{eff}$.
The present value of the deceleration parameter is
\begin{equation}
\frac{\ddot{a}}{aH^{2}}\mid _{0} =  - \frac{1}{2}\left\{1 + 3w\Omega_{x_{0}}\right\}
 \ ,
\label{q0x}
\end{equation}
where $\Omega_{x_0} = \frac{8\pi G \rho_{x_0}}{3 H_{0}^{2}}$.

To have a viable cosmological model of transient acceleration, formula (\ref{accOmega2}) should admit a transition from $\frac{\ddot a}{a} <0$ to $\frac{\ddot a}{a} >0$ before the present time, i.e., for $a < 1$. If, moreover, the accelerated expansion is a transient phenomenon, there should be a change back  from $\frac{\ddot a}{a} >0$ to $\frac{\ddot a}{a} <0$ at some time, which may be close to the present epoch or at a future period $a > 1$.
In the expression (\ref{accOmega2}) the $a^{-3}$ terms on the right hand side dominate for small values of $a$, i.e., there is decelerated expansion for $a\ll 1$ provided the condition
\begin{equation}
\label{dec2}
\Omega_{m_{0}}  > \bar{K}\exp (-1/\sigma^{2})
\end{equation}
is satisfied.
This condition puts an upper limit on the admissible interaction strength. In the non-interacting limit it just expresses  the positivity of the matter energy density.
Let's consider now the case $a\gg 1$. The dominating contribution in the braces on the right-hand side  of
(\ref{accOmega2}) are then  given by the constant term $-2\rho_{x_0}^{eff}$.
As long as $\rho_{x_0}^{eff} > 0$, however small it may be, we will have $\frac{\ddot a}{a} >0$ for $a\gg 1$,
i.e., there is no transition back to decelerated expansion.
This holds, in particular, in the non-interacting limit which reproduces the $\Lambda$CDM model. Then
$\Omega_{x_0}^{eff}$ reduces to $\Omega_{x_0}$, equivalent to $\Omega_{\Lambda_0}$. For $a\gg 1$ this term will always dominate the dynamics.
An obvious way to obtain decelerated expansion for $a\gg 1$ is to
put $\rho_{x_0}^{eff} = 0$  in
(\ref{accOmega2}). This corresponds to a vanishing \textit{total} cosmological constant.
In other words, part of the interaction cancels the ``bare" cosmological constant, described by $\rho_{x_0}$.
Under this condition it is exclusively the remaining part of the interaction which potentially can trigger a period of accelerated expansion.
In such a case one  obtains
from (\ref{rh2}) that
\begin{equation}
\Omega_{x_{0}} = \bar{K}\,\exp (-1/\sigma^{2})
\left[\frac{3}{2}\sigma^{2} - 1\right]\ .
\label{r20o}
\end{equation}
Then the energy densities of the dark components are (\ref{rhom2}) for $\rho_{m}$
and (\ref{rh2}) for $\rho_{x}$ with $\rho_{x0}^{eff} = 0$, i.e.,
\begin{equation}
\rho_{x} =
\frac{3}{2}\sigma^{2} K\,\exp\left(-a^{2}/\sigma^{2}\right)
\left(1 - \frac{2}{3}\frac{a^{2}}{\sigma^{2}}\right)
\ .
\label{rh2mod}
\end{equation}
Notice that for $K>0$ a positive value of $\rho_{x}$ requires $a<\sqrt{\frac{3}{2}} \sigma$.
For any $a> \sigma$ the entire quantity (\ref{rh2mod}) is exponentially suppressed,
the amount of $\rho_{x}$ tends to zero (possibly through an intermediate period with $\rho_{x}<0$).
With $\rho_{m}$ from (\ref{rhom2}) and
$\rho_{x }$ from (\ref{rh2mod}),
the Hubble rate for this model is
\begin{equation}
\label{H2}
\frac{H^{2}}{H_{0}^{2}} = \frac{1 - \frac{3}{2} \sigma^{2}\bar{K}\exp
(-1/\sigma^{2})}{a^{3}} + \frac{3}{2} \sigma^{2}\bar{K} \exp
(-a^{2}/\sigma^{2})
\ .
\end{equation}
Both for $a\ll 1$ and for $a\gg 1$ one has
\begin{equation}
\label{H><}
\frac{H^{2}}{H_{0}^{2}} \approx \frac{1 - \frac{3}{2} \sigma^{2}\bar{K}\exp
(-1/\sigma^{2})}{a^{3}} \qquad\qquad (a\ll 1, \quad a\gg 1)
\ .
\end{equation}

\begin{figure}[h]
\includegraphics[scale=0.50]{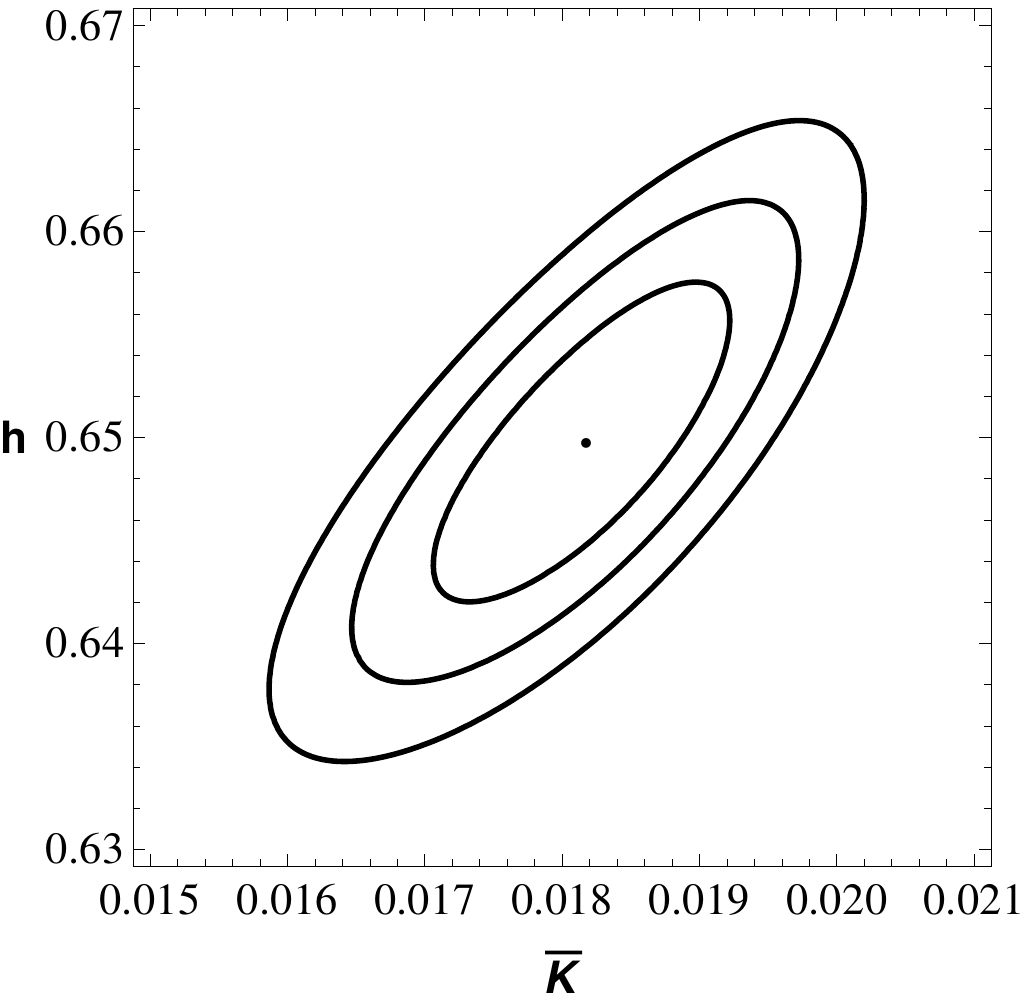}
\includegraphics[scale=0.53]{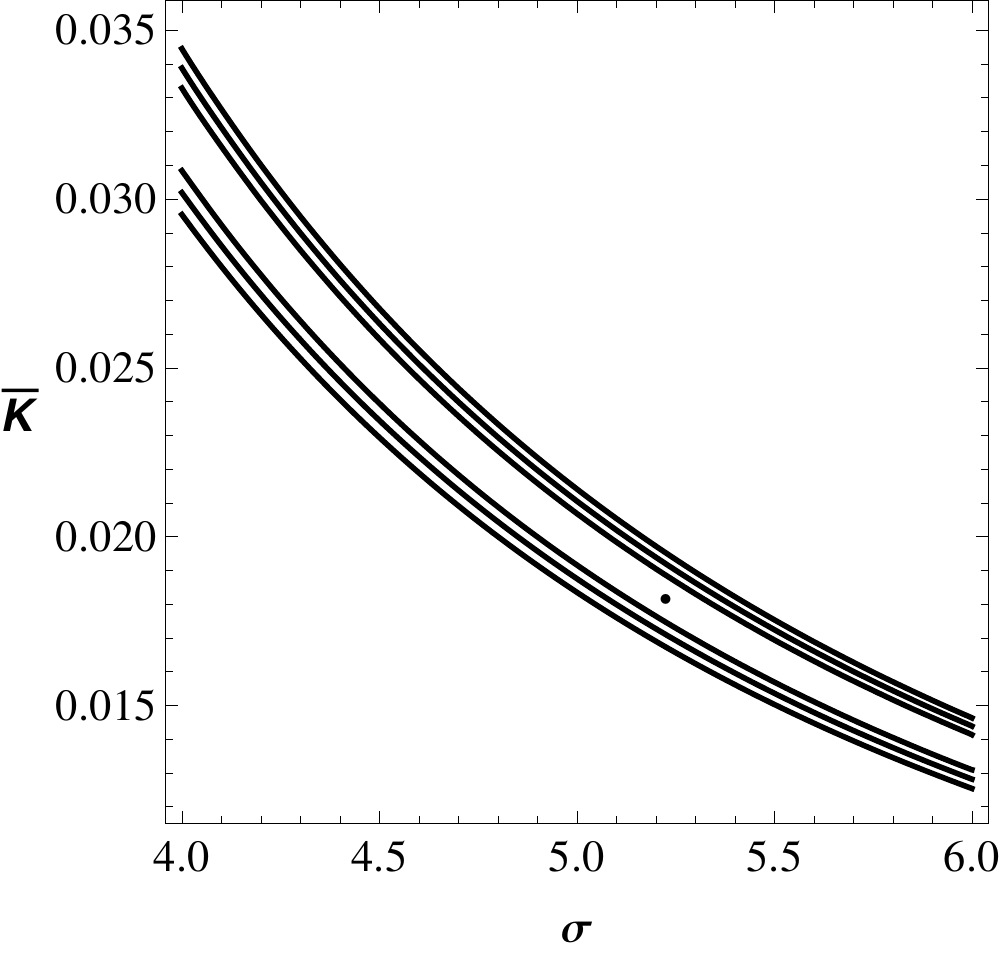}
\includegraphics[scale=0.50]{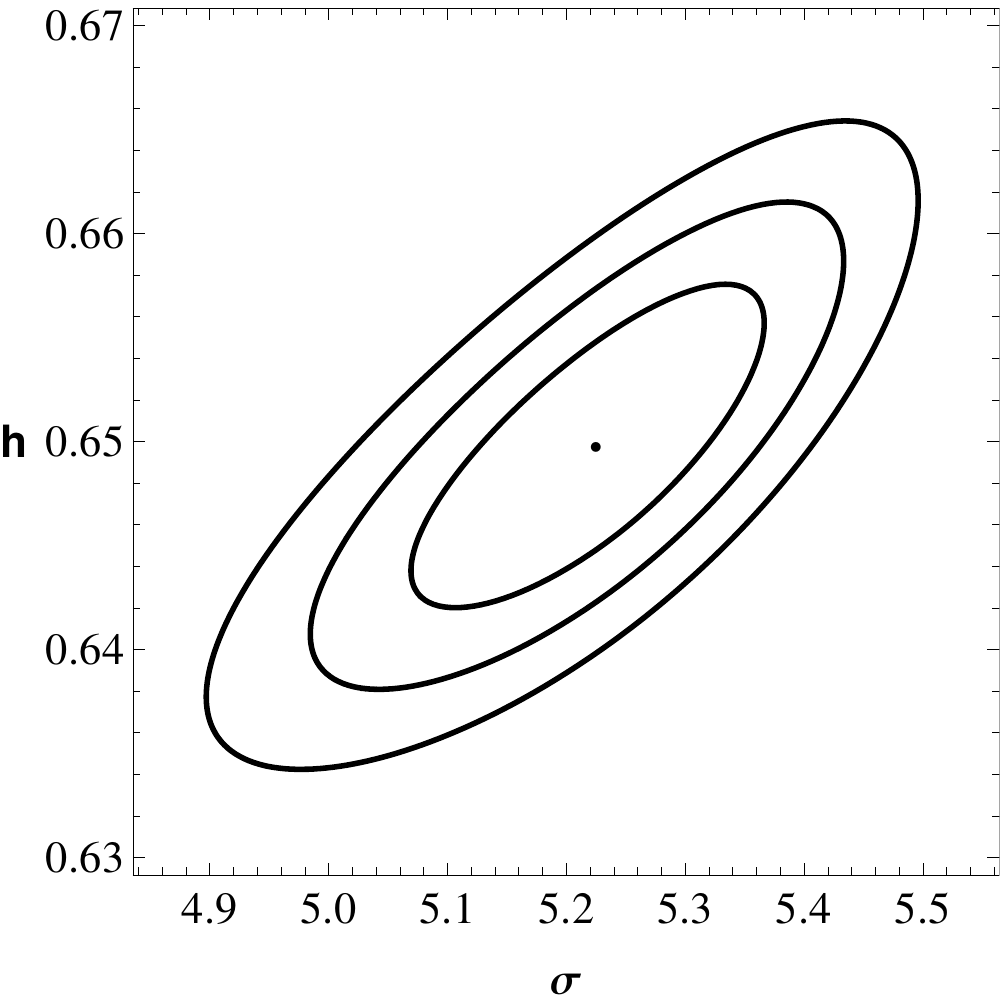}
\caption{Two-dimensional probability contours ($1\sigma$, $2\sigma$ and $3\sigma$), based on the
Constitution data, for all combinations of the parameters $h$, $\bar{K}$ and $\sigma$.}
\label{contourconstitution}
\end{figure}

\begin{table}[!t]
\begin{center}
\begin{tabular}{|c|c|c|c|@{\vrule height 11pt depth 5pt width 0pt}} \hline
$\chi^{2}_{\mbox{\tiny{min}}}$ & $\sigma$ & $\bar{K}$ & $h$\\ \hline
\hline
$465.5$ & $5.23 ^{+0.05}_{-0.05}$   & $0.018 ^{+0.0004}_{-0.0004} $  &  $0.65 ^{+0.003}_{-0.003}$ \\ \hline
\end{tabular}
\end{center}
\caption{Best-fit values, based on the
Constitution data,  for the parameters $h$, $\bar{K}$ and $\sigma$.}
\label{tabconstitution}
\end{table}

The acceleration equation
becomes
\begin{equation}
\label{accOmega2fin}
\frac{\ddot a}{a}=-\frac{1}{2}H_{0}^{2}
\left\{\frac{1 - \frac{3}{2}\bar{K}\sigma^{2}\exp
(-1/\sigma^{2})}{a^{3}}
+ 3 \bar{K} \exp
(-a^{2}/\sigma^{2}) \left[a^{2} - \sigma^{2}\right]\right\} \ .
\end{equation}
To have decelerated expansion for $a\ll 1$,
\begin{equation}
\label{condll2}
 \bar{K} \,\sigma^{2}\,\exp
(-1/\sigma^{2}) < \frac{2}{3}
\end{equation}
has to be required.
This condition is similar to (\ref{dec2}).
The zeros of (\ref{accOmega2fin}) determine the values $a_{q}$ of $a$ at which transitions between decelerated and accelerated expansion (or the reverse) occur, namely
\begin{equation}
\frac{3}{2}\sigma^{2} \bar{K}\,\exp (-1/\sigma^{2}) +
3 \bar{K} a_{q}^3\,\exp
(-a_{q}^2/\sigma^2) \left[\sigma^{2} - a_{q}^2\right]
 = 1
\ .
\label{consist2}
\end{equation}
The condition to have acceleration at the present epoch with $a=1$ is
\begin{equation}
\frac{\ddot{a}}{aH^{2}}\mid _{0} \ > 0 \quad \Leftrightarrow \quad
\bar{K}\,\exp (-1/\sigma^{2})\left[\sigma^{2} - \frac{2}{3}\right] > \frac{2}{9} \ .
\label{condacc1}
\end{equation}
If the inequality (\ref{condacc1}) holds, we may have present acceleration under the condition $\rho_{x_0}^{eff}=0$, i.e., a vanishing total cosmological constant.
Obviously, the normalized interaction strength  $\bar{K}$ has to be larger than a threshold value to realize this configuration. The condition (\ref{condacc1}) is consistent with (\ref{q0x}) if the latter is combined with (\ref{r20o}).
On the other hand, we have the upper limit (\ref{condll2}). This means, there exists a range for admissible values of the interaction strength, determined by
\begin{equation}
\frac{2}{9}\frac{e^{1/\sigma^{2}}}{\sigma^{2} - \frac{2}{3}} < \bar{K} <
\frac{2 e^{1/\sigma^{2}}}{3\sigma^{2}}\ .
\label{region2}
\end{equation}
The parameters $\bar{K}$ and $\sigma^{2}$ enter the present ratio of the energy densities for which we find
\begin{equation}
\frac{\rho_{x0}}{\rho_{m0}} =
\frac{\bar{K}\exp(-1/\sigma^{2})\left(\frac{3}{2}\sigma^{2} - 1\right)}{1 - \bar{K}\exp(-1/\sigma^{2})\left(\frac{3}{2}\sigma^{2} - 1\right)}
\ .
\label{r0}
\end{equation}
The results of a Bayesian statistical analysis, using the 397 SNIa data of the Constitution sample \cite{Hicken}, are shown in Fig.~\ref{contourconstitution} and in Table~\ref{tabconstitution}.
For the best-fit values in Table~\ref{tabconstitution} the inequalities (\ref{region2}) are satisfied and for the ratio (\ref{r0}) we obtain $\frac{\rho _{x_{0}}}{\rho _{m_{0}}} =  2.275 ^{+0.355}_{-0.285}$.
The range in (\ref{region2}) specifies to
$0.0086 < \bar{K} < 0.0253$.
(For the integration a range $-0.1 < \bar{K} < 0.1$ was used but robustness tests showed that the results are independent of this choice.)
The numerical value for the left-hand side in (\ref{condll2}) becomes $0.475 < \frac{2}{3}$ and for the right-hand side of (\ref{condacc1}) we find $0.463 > \frac{2}{9}$.

The relation (\ref{accOmega2fin}) may be compared with the corresponding expression of the
$\Lambda$CDM model:
\begin{equation}
\label{acclcdm}
\frac{\ddot a}{a}|_{\Lambda CDM}=-\frac{1}{2}H_{0}^{2}
\left\{\frac{1 - \Omega_{\Lambda}}{a^{3}}
- 2 \Omega_{\Lambda}\right\} \ .
\end{equation}
The interaction term
in (\ref{accOmega2fin}) plays the role of $\Omega_{\Lambda}$ in (\ref{acclcdm}).
Now we know that the $\Lambda$CDM model provides a fairly good description of the present universe, i.e., for $a=1$. This suggest positive values of the interaction constant $\bar{K}$ together with $\sigma >1$, which is indeed confirmed by our analysis.

The relation between $\rho_{m_0}$ in (\ref{rhom2}) and $\tilde\rho_{m_0}$ in the definition of $K$ following
(\ref{rhom2}) is
\begin{equation}
\tilde\rho_{m_0} = \frac{\rho_{m_0}}{1 + c\,\exp\left(-1/\sigma^{2}\right)}
\ .
\label{rm0rm0t}
\end{equation}
With the help of the definition $\Omega_{m_0} = \frac{8\pi G \rho_{m_0}}{3 H_{0}^{2}}$, the constant $c$ may be written as
\begin{equation}
c = \frac{\bar{K}}{\Omega_{m_0} - \bar{K}\,\exp\left(-1/\sigma^{2}\right)}
\ .
\label{c}
\end{equation}
The quantity $g$ in (\ref{g}) is then given by
\begin{equation}\label{dotg}
g
= \frac{\bar{K}\,a^{5}\,\exp\left(-a^{2}/\sigma^{2}\right)}{\Omega_{m_0} - \bar{K}\,\exp\left(-1/\sigma^{2}\right)}
= \frac{\bar{K}\,a^{5}\,\exp\left(-a^{2}/\sigma^{2}\right)}{1 - \frac{3}{2}\sigma^{2}\bar{K}\,\exp\left(-1/\sigma^{2}\right)} \ ,
\end{equation}
 and the interaction term $\frac{Q}{\rho_{m}} = \frac{\dot{g}}{1+g}$ becomes
\begin{equation}
\frac{Q}{\rho_{m}} = \bar{K}\,H\,
\frac{\left(5 - 2\frac{a^{2}}{\sigma^{2}}\right)a^{5}
\,\exp\left(-a^{2}/\sigma^{2}\right)}{\Omega_{m_0} - \bar{K}\left[\exp\left(-1/\sigma^{2}\right) -
a^{5}\,\exp\left(-a^{2}/\sigma^{2}\right)\right]} = \frac{g}{1+g}\left(5 - 2\frac{a^{2}}{\sigma^{2}}\right)\,H
\ .
\label{Q/rhom}
\end{equation}
It is obvious that a transfer of energy from dark energy to dark matter, characterized by $Q>0$, requires $a^{2}<\frac{5}{2}\sigma^{2}$.
While positive values of $Q$ seem to be favored on thermodynamical grounds \cite{diegowang}, the observational situation is less clear \cite{royM}. Moreover, negative values of both the energy density (\ref{rh2mod})
and of $Q$ in (\ref{Q/rhom}) for sufficiently large values of the scale factor are exponentially suppressed and the total energy remains always positive.
According to this scenario dark energy is transformed into dark matter in the past, at present and in the future until $a< \sqrt{\frac{5}{2}}\sigma \approx 8$. At $a= \sqrt{\frac{3}{2}}\sigma \approx 6$ the density of the dark energy becomes negative. However, it decays only until $a \approx 8$. At this point the direction of the energy transfer reverses and for $a \gg \sigma$ it tends to zero exponentially.

\begin{figure}[!h]
\centering
\includegraphics[width=1.00\textwidth]{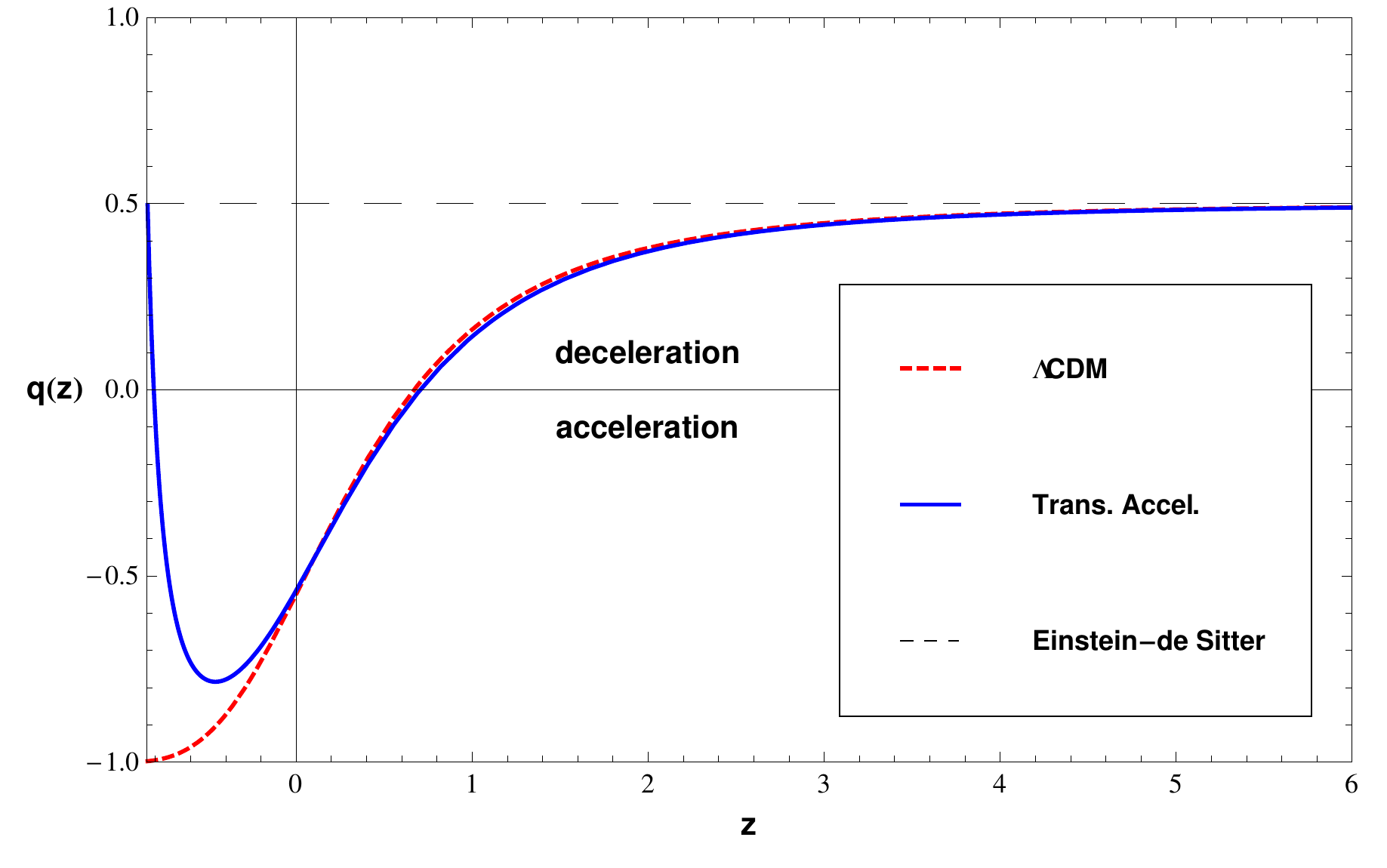}
\caption{The deceleration parameter of the transient acceleration model as function of the redshift
for the best-fit parameters in Table~\ref{tabconstitution} (solid line). The dashed line shows the corresponding dependence for the $\Lambda$CDM model. The value $q= \frac{1}{2}$ corresponds to the Einstein-de Sitter universe.}
\label{figq}
\end{figure}

The background interaction term $Q$ may also be written as
\begin{equation}
Q = 3 \mu H \rho_{x} = \mu \Theta \rho_{x}
\label{Q2rho}
\end{equation}
with
\begin{equation}
\mu\left(a\right) = \frac{2}{9} \frac{5 - 2\frac{a^{2}}{\sigma^{2}}}
{\frac{\sigma^{2}}{a^{2}} - \frac{2}{3}}
\ .
\label{mu}
\end{equation}
The consistency of the model can be checked by realizing $\dot{\rho}_{x} = -Q$ (cf.Eq.~(\ref{dotrhox})) for $w=-1$) with $\rho_{x}$ from (\ref{rh2mod}) and
$Q$ from (\ref{Q/rhom}) with $\rho_{m}$ from (\ref{rm}) with $g$ from (\ref{g}).
The behavior of the deceleration parameter for the best-fit values of Table~\ref{tabconstitution} is shown in Fig.~\ref{figq}. According to our model, the Universe is still in accelerated expansion at the present epoch but $q(z)$ will go through a minimum in the future and enter a phase of decelerated expansion again. In the far-future limit $a\gg 1$ (not shown in the figure) $q$ will approach $q=\frac{1}{2}$ again.

\section{Newtonian perturbation theory}
\label{newton}

In this section we focus on perturbations with wavelengths much smaller than the Hubble radius.
Under this condition, the dynamics is well approximated by a Newtonian analysis.
Afterwards we shall clarify how this approximation fits into a general relativistic scheme.
Starting point for a Newtonian treatment is the matter energy balance equation $\frac{\partial\rho_{m}}{\partial t} + \left(\rho_{m}v^{\alpha}\right)_{,\alpha} = Q$, where greek indices run over $1$, $2$ and $3$ and $v^{\alpha}$ is the (non-relativistic) matter velocity.
The perturbed energy balance is, in first order,
\begin{equation}
\dot{\hat{\rho}}_{m} + 3 \frac{\dot{a}}{a}\hat{\rho}_{m}
+ \rho_{m} \hat{v}^{\alpha}_{,\alpha}
= \hat{Q}\ .
\label{contpert2}
\end{equation}
Here, a hat on top of the symbol denotes the first-order perturbation of the corresponding quantity.
It is convenient to introduce the fractional quantity $\delta_{m}\equiv\frac{\hat{\rho}_{m}}{\rho_{m}}$.
Eq.~(\ref{contpert2}) is then equivalent to
\begin{equation}
\dot{\delta}_{m} + \hat{v}^{\alpha}_{,\alpha} = \frac{1}{\rho_{m}}\left(\hat{Q} - Q\delta_{m}\right) \ .
\label{ddeltam}
\end{equation}
The right-hand side of this equation describes the influence of the interaction on the perturbation dynamics.
In the interaction-free limit it reduces to zero. Since the Newtonian model does not specify $\hat{Q}$ we shall assume for simplicity $\hat{Q} = \beta Q\delta_{m}$ where $\beta$ is a constant. In the subsequent section we shall look at this term more carefully. The limit $\beta = 0$ corresponds to an interaction only in the background. For $\beta = 1$ the interaction does not affect the perturbation dynamics. Eq.~(\ref{ddeltam}) specifies to
\begin{equation}
\dot{\delta}_{m} + \hat{v}^{\alpha}_{,\alpha} = - \left(1 - \beta\right)\frac{Q}{\rho_{m}}\delta_{m} \ .
\label{dotdelta}
\end{equation}
Assuming also, that there is separate momentum conservation of both components (the more general case of a coupling also via exchange of momentum will be considered in the following section),
the non-relativistic Euler equation for the matter reads $\frac{\partial v_{\alpha}}{\partial t} + \left(v^{\mu}\nabla_{\mu}
\right)v_{\alpha} = - \phi_{,\alpha}$, where $\phi$ is the gravitational potential.
From the first-order Euler equation we find
\begin{equation}
\dot{\hat{v}}_{\alpha} = -\frac{\dot{a}}{a}\hat{v}_{\alpha} - \hat{\phi}_{,\alpha}\ .
\label{euler}
\end{equation}
Introducing comoving coordinate $q^{\alpha}$ by $x^{\alpha} = a q^{\alpha}$, differentiating (\ref{euler}) with respect to $q^{\alpha}$ and combing the result with (\ref{dotdelta}) results in
\begin{equation}
\ddot{\delta}_{m} + \left(2H + \left(1 - \beta\right)\frac{Q}{\rho_{m}}\right) \dot{\delta}_{m}
+ \left(1 - \beta\right)\left[2H\frac{Q}{\rho_{m}} + \left(\frac{Q}{\rho_m}\right)^{\displaystyle\cdot}\right]\delta_{m}
- \frac{1}{a^{2}}\Delta_{q}\hat{\phi} = 0
\ ,
\label{dddm}
\end{equation}
where $\Delta_{q}$ is the Laplacian with respect to the comoving coordinates.
Equation (\ref{dddm}) demonstrates the influence of the interaction on the perturbation dynamics. Both the coefficients of
$\dot{\delta}_{m}$ and $\delta_{m}$ depend on $Q$ explicitly. Even for $\beta = 1$, the case in which the interaction is not directly felt at the perturbative level, the Hubble rate $H$ is essentially determined by the interaction according to (\ref{H2}).
The first-order field equation of Newtonian gravity reads
\begin{equation}
\frac{1}{a^{2}}\Delta_{q}\hat{\phi} = 4\pi G\left(\delta\rho_{m} + \delta\rho_{x}\right) =
4\pi G\left(\rho_{m} \delta_{m} + \rho_{x}\delta_{x}\right)
\ ,
\label{Nfield}
\end{equation}
where $\delta_{x} \equiv \frac{\hat{\rho}_{x}}{\rho_{x}}$. In many studies of the growth rate of matter perturbations the dark-energy perturbations are neglected. This corresponds to assuming $\delta_{x} = 0$
in (\ref{Nfield}). However, this is strictly justified only for a cosmological constant. In dynamical dark-energy models $\delta_{x}$ is different from zero and the matter perturbations are coupled
to the dark-energy perturbation. Neglecting this influence may result in an incorrect interpretation
of observational data \cite{Park-Hwang}. On the other hand, for specific models the coupling can indeed be shown to be
negligible on small scales \cite{saulo}. In order to obtain a closed second-order equation for $\delta_{m}$ we shall assume here
a simple relation of proportionality $\delta_{x} = \alpha\delta_{m}$ between  $\delta_{x}$ and $\delta_{m}$, where $\alpha$ is constant (cf. \cite{ivan}). The limit $\alpha = 0$ corresponds to vanishing dark-energy fluctuations. For any $\alpha$ of the order
of one, the dark-energy perturbations are relevant for structure formation.
Under this condition we have
\begin{equation}
\frac{1}{a^{2}}\Delta_{q}\hat{\phi} = 4\pi G\left(\rho_{m} + \alpha\rho_{x}\right)\delta_{m}
= \frac{3}{2}H^{2}\left[\left(1-\alpha\right)\frac{\rho_{m}}{\rho} + \alpha\right]\delta_{m}
\
\label{}
\end{equation}
for the last term in Eq.~(\ref{dddm}).
For the term that multiplies $\delta_{m}$ in (\ref{dddm}) one finds
\begin{equation}\label{dQrm}
\left(\frac{Q}{\rho_m}\right)^{\displaystyle\cdot} + 2 H\,\frac{Q}{\rho_{m}}
= H\left[\frac{1}{2} + A + B - 4\frac{\frac{a^{2}}{\sigma^{2}}}{5 - 2\frac{a^{2}}{\sigma^{2}}}\right]
\,\frac{Q}{\rho_{m}}  \ ,
\end{equation}
where
\begin{equation}\label{A}
A(a) \equiv \frac{3}{2a} + \frac{\dot{H}}{aH^{2}} =
\frac{9}{4}\sigma^{2} \bar{K}\exp\left(-a^{2}/\sigma^{2}\right)\left(1 - \frac{2}{3}\frac{a^{2}}{\sigma^{2}}\right)\,\frac{H^{2}_{0}}{H^{2}}
\ ,
\end{equation}
with
\begin{equation}\label{H0/H}
\frac{H_{0}^{2}}{H^{2}} = \frac{a^{3}}{
1 - \frac{3}{2}\sigma^{2}\bar{K}\left[\exp\left(-1/\sigma^{2}\right)
- a^{3}\exp\left(-a^{2}/\sigma^{2}\right)\right]}
\
\end{equation}
and
\begin{equation}\label{B}
B(a) \equiv \frac{5 - 2\frac{a^{2}}{\sigma^{2}}}{1+g}
\ .
\end{equation}
With the help of these relations and abbreviations, the basic perturbation equation (\ref{dddm})
is written as
\begin{equation}\label{dddmeff}
\ddot{\delta}_{m} + \left[2 + \left(1-\beta\right)gB\right]H\dot{\delta}_{m}
- 4\pi G_{eff}\rho_{m}\delta_{m} = 0\ ,
\end{equation}
with an effective gravitational constant
\begin{equation}\label{Geff}
G_{eff} \equiv G\left\{1 + \alpha\frac{1 - \Omega_{m}}{\Omega_{m}} - \frac{2}{3\Omega_{m}}\left(1-\beta\right)\frac{g}{1+g}
\left[\left(5-2\frac{a^{2}}{\sigma^{2}}\right)\left(\frac{1}{2}+A+B\right) -
4 \frac{a^{2}}{\sigma^{2}}\right]\right\}\ .
\end{equation}
Recall, that the interaction also explicitly enters the ``friction" term in addition to its influence
on the Hubble rate itself. This additional term in the factor that multiplies $\dot{\delta}_{m}$
in (\ref{dddmeff}) vanishes only for $\beta = 1$.
Changing to $a$ as independent variable,
\begin{equation}
\dot{\delta}_{m} = \delta_{m}^{\prime}aH \ , \qquad \ddot{\delta}_{m} = a^{2}H^{2}
\left[\delta_{m}^{\prime\prime} + \frac{1}{a}\delta_{m}^{\prime} + \frac{H^{\prime}}{H}\delta_{m}^{\prime}\right]\ ,
\label{prime}
\end{equation}
where $\delta_{m}^{\prime} \equiv \frac{d \delta_{m}}{d a}$, we obtain the final equation
\begin{equation}
\delta_{m}^{\prime\prime} + U(a)\delta_{m}^{\prime} + V(a) \delta_{m} = 0
\ ,
\label{deltaprpr}
\end{equation}
with
\begin{equation}
U(a)= \frac{1}{a}\left[\frac{3}{2} + A(a)
+ \left(1 - \beta\right)g(a)B(a)\right]\ , \qquad A = \frac{3}{2a} + \frac{H^{\prime}}{H}
\
\label{U}
\end{equation}
and
\begin{equation}
V(a)= -\frac{3}{2a^{2}}\Omega_{m}\frac{G_{eff}}{G}
\ ,
\label{}
\end{equation}
where
\begin{equation}\label{}
\Omega_{m} = \left\{\Omega_{m0} +  \bar{K} \left[a^{5}\exp(-a^{2}/\sigma^{2}) - \exp(-1/\sigma^{2})\right]\right\}\frac{1}{a^{3}}\frac{H_{0}^{2}}{H^{2}}
\ .
\end{equation}
With $\Omega_{m0} = 1 -  \bar{K}\exp(-1/\sigma^{2})\left(\frac{3}{2}\sigma^{2} -1\right)$ one realizes that
$\Omega_{m} \approx 1$ both for high redshifts $a\ll 1$ and in the long-time limit $a\gg 1$.

\begin{figure}[h]
\includegraphics[scale=0.68]{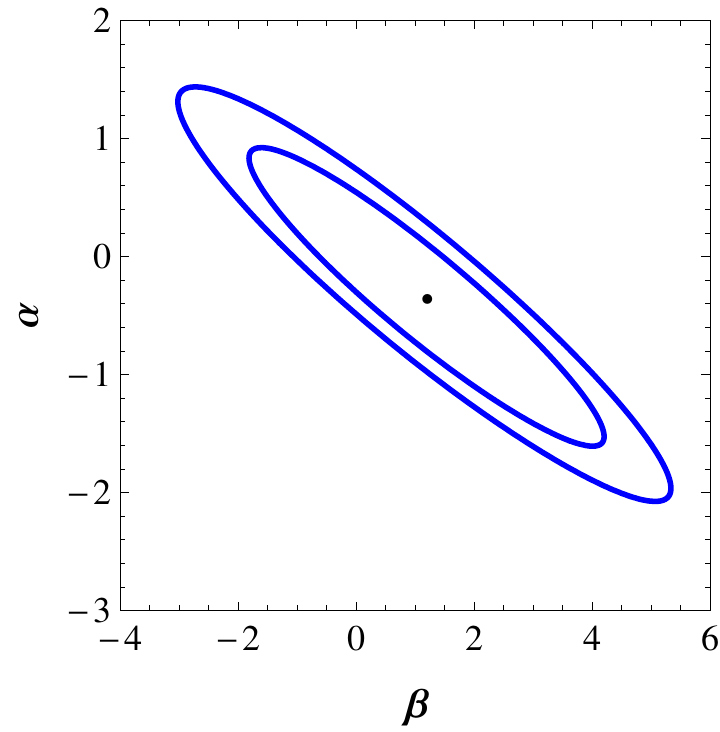}
\includegraphics[scale=0.68]{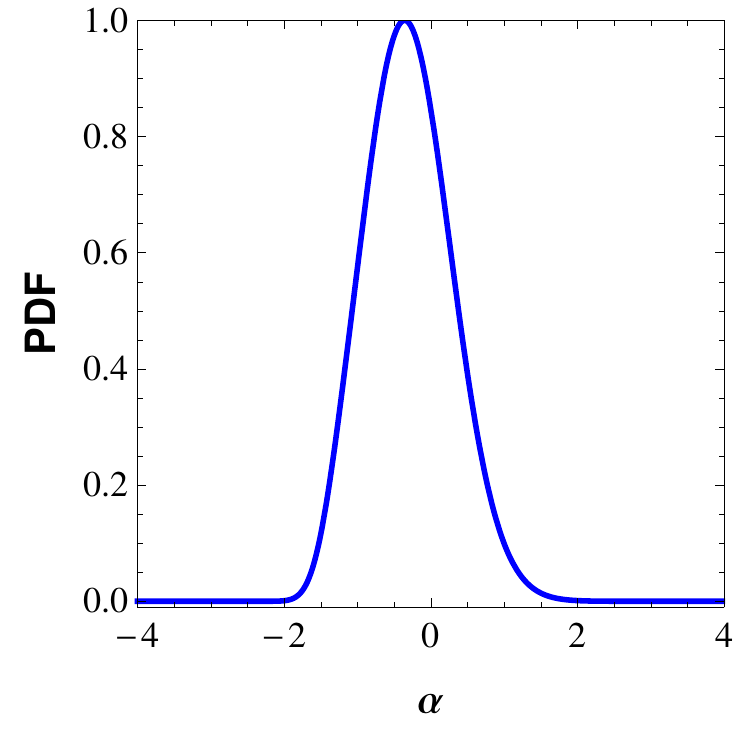}
\includegraphics[scale=0.68]{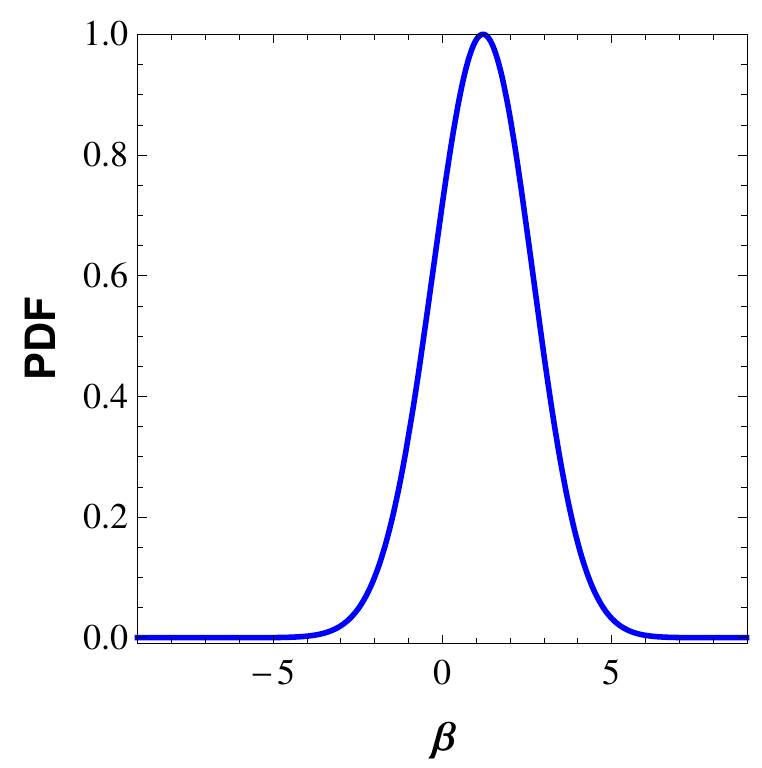}
\caption{Statistical analysis for the parameters $\alpha$ and $\beta$, based on the  grow-rate data in \cite{gong} and \cite{blake}.
Left panel: contour plot for the best-fit values of Table~\ref{tabalphabeta}. Center and right panels: probability distribution functions (PDFs) for $\alpha$ and $\beta$, respectively.}
\label{figalphabeta}
\end{figure}

\begin{figure}[h]
\includegraphics[scale=0.57]{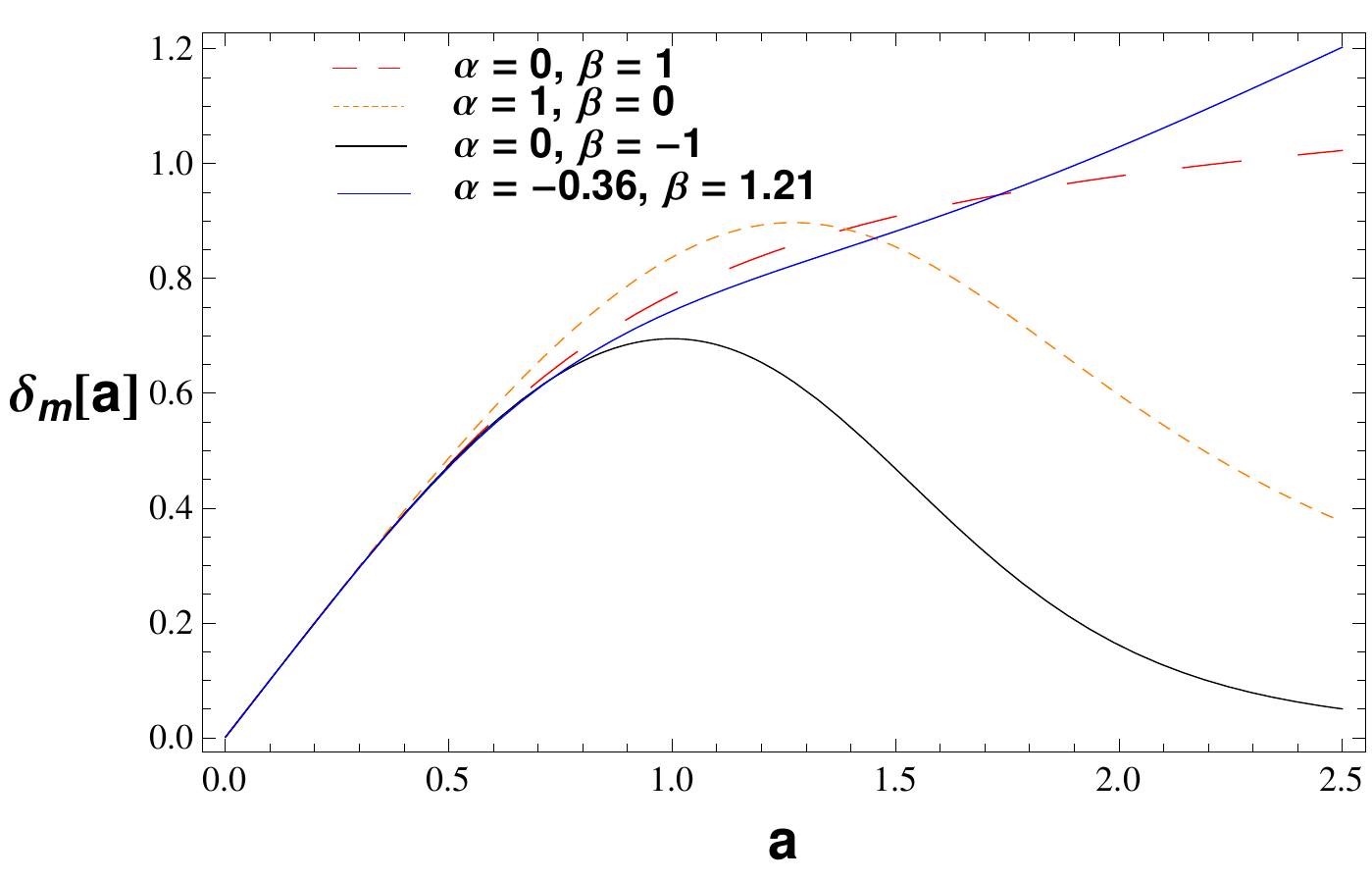}
\includegraphics[scale=0.53]{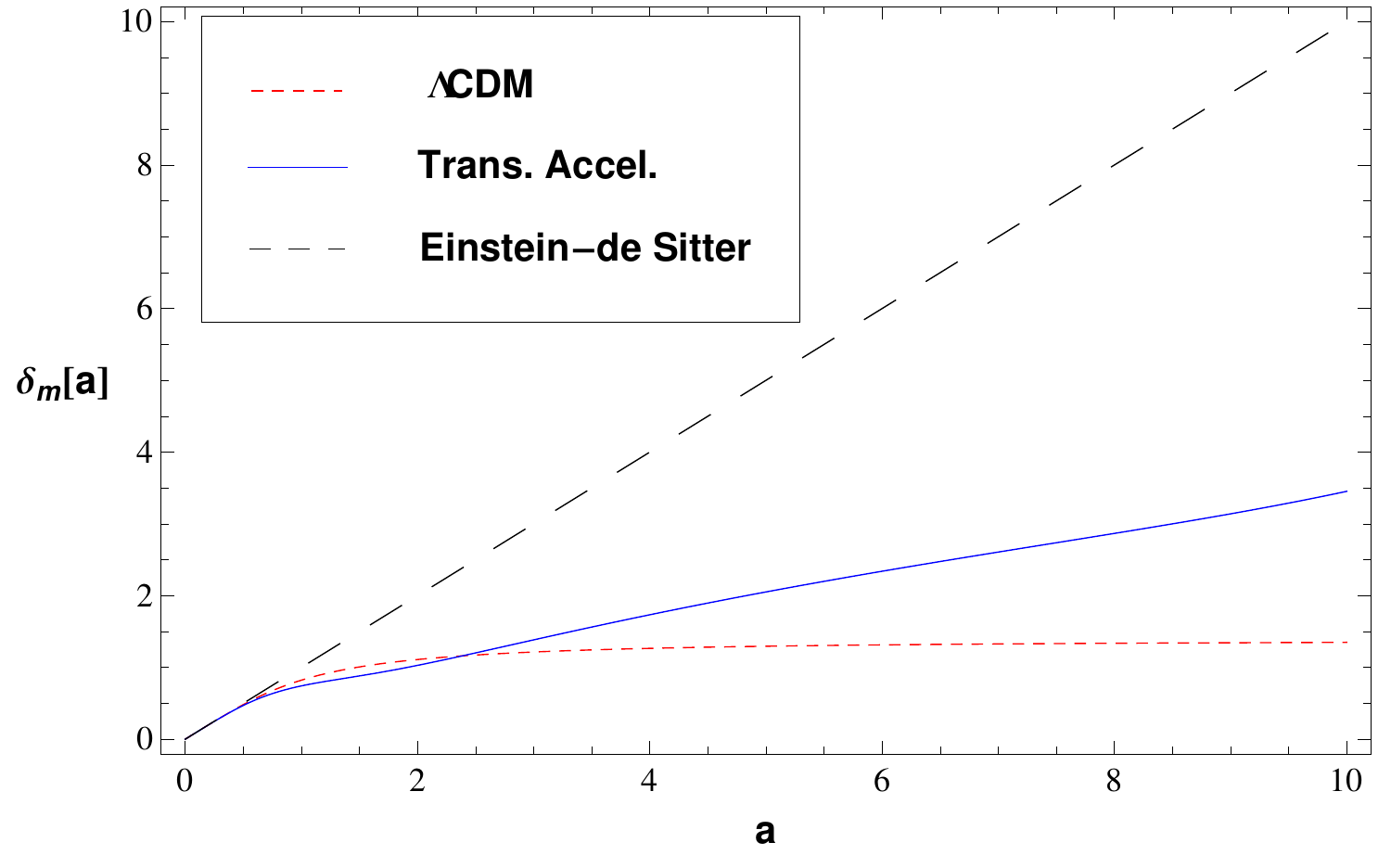}
\caption{Fractional density perturbation as a function of the scale factor. Left panel: perturbation behavior for different values of $\alpha$ and $\beta$. Right panel: comparison between our  model
with the best-fit parameters of Tables~\ref{tabconstitution} and \ref{tabalphabeta} and the $\Lambda$CDM model. The straight line shows the corresponding increase for the Einstein-de Sitter model.}
\label{figdelta1}
\end{figure}

Equation (\ref{deltaprpr}) is the central equation for the Newtonian perturbation analysis. The free parameters are $H_{0}$,  $\bar{K}$ and $\sigma$ as well as $\alpha$ and $\beta$. In the non-interacting limit and
for the one-component case $\rho = \rho_{m}$
one recovers the perturbation equation
\begin{equation}
\delta_{m}^{\prime\prime} + \frac{3}{2a}\delta_{m}^{\prime} - \frac{3}{2a^{2}} \delta_{m} = 0
\
\label{deltaprprni}
\end{equation}
for the Einstein-de Sitter universe.
 Moreover, (\ref{deltaprprni}) is also the limit of (\ref{deltaprpr}) both for $a\ll 1$ and for $a\gg 1$, since all the interaction terms are vanishing under these conditions.
 In general, the interaction influences the perturbation dynamics through all the $\bar{K}$ terms in the coefficients $U(a)$ and $V(a)$ in (\ref{deltaprpr}) which also includes the modification of the Hubble rate due to the interaction on the background level.
 To test this model, we performed a Bayesian statistical analysis, based on the growth rate data collected in  \cite{gong} and those of the WiggleZ survey \cite{blake}.
 The left panel of Fig.~\ref{figdelta1} shows the dependence of $\delta_{m}$ for different values for $\alpha$ and $\beta$. The thick solid (blue) curve corresponds to the best-fit values of Table~\ref{tabalphabeta}, visualized in Fig.~\ref{figalphabeta}.
The right panel compares the best-fit curve with the corresponding behavior for an Einstein-de Sitter universe and with the $\Lambda$CDM model. The growth of the matter perturbations is reduced compared
with its $\Lambda$CDM counterpart for values $a\approx1$. For larger values of the scale factor the perturbations grow again, whereas they stay constant for the $\Lambda$CDM model.
If the recent WiggleZ data are not included, there exists a maximum in $\delta_{m}(a)$ close to the present time, similar to some of the curves in the left panel.


\begin{table}
\begin{center}
\begin{tabular}{|c|c|c|@{\vrule height 11pt depth 5pt width 0pt}} \hline
$\chi^{2}_{\mbox{\tiny{min}}}$ & $\alpha$ & $\beta$ \\ \hline
\hline
$7.457$ & $-0.357^{+1.057}_{-1.403}$   & $ 1.205^{+2.504}_{-2.504} $   \\ \hline
\end{tabular}
\end{center}
\caption{Best-fit values for the parameters $\alpha$ and $\beta$, based on the growth-rate data in \cite{gong} and \cite{blake}.}
\label{tabalphabeta}
\end{table}

 The different growth of matter perturbations has been used in the literature primarily to discriminate between a GR-based behavior and alternative theories of gravity \cite{linderjenkins,uzan,hutererlinder,diportoamendola,lahav,mantz,nesserisperi,polarski,gong,gongishak,dosset}. However, also the impact of interactions on the perturbation growth has been investigated \cite{maartensgrowth,ivan,saulo}.
It is convenient to introduce the growth  rate
\begin{equation}
 f:= \frac{d \ln \delta_{m}}{d \ln a} \, , \label{gfunction}
\end{equation}
in terms of which the basic equation (\ref{deltaprpr}) takes the form
\begin{equation}
\frac{df}{d\ln a}\,  + \, f^{2} +  \left[a U(a) - 1\right]f + a^{2}\,V(a)= 0
 \, .
\label{eqf}
\end{equation}
The last equation can also be written as
\begin{equation}
\frac{df}{d\ln a}\,  + \, f^{2} +  \left[a U(a) - 1\right]f = \frac{3}{2}\frac{G_{eff}}{G}\Omega_{m}
 \,.
\label{eqfGeff}
\end{equation}

\begin{figure}[h]
\includegraphics[scale=0.49]{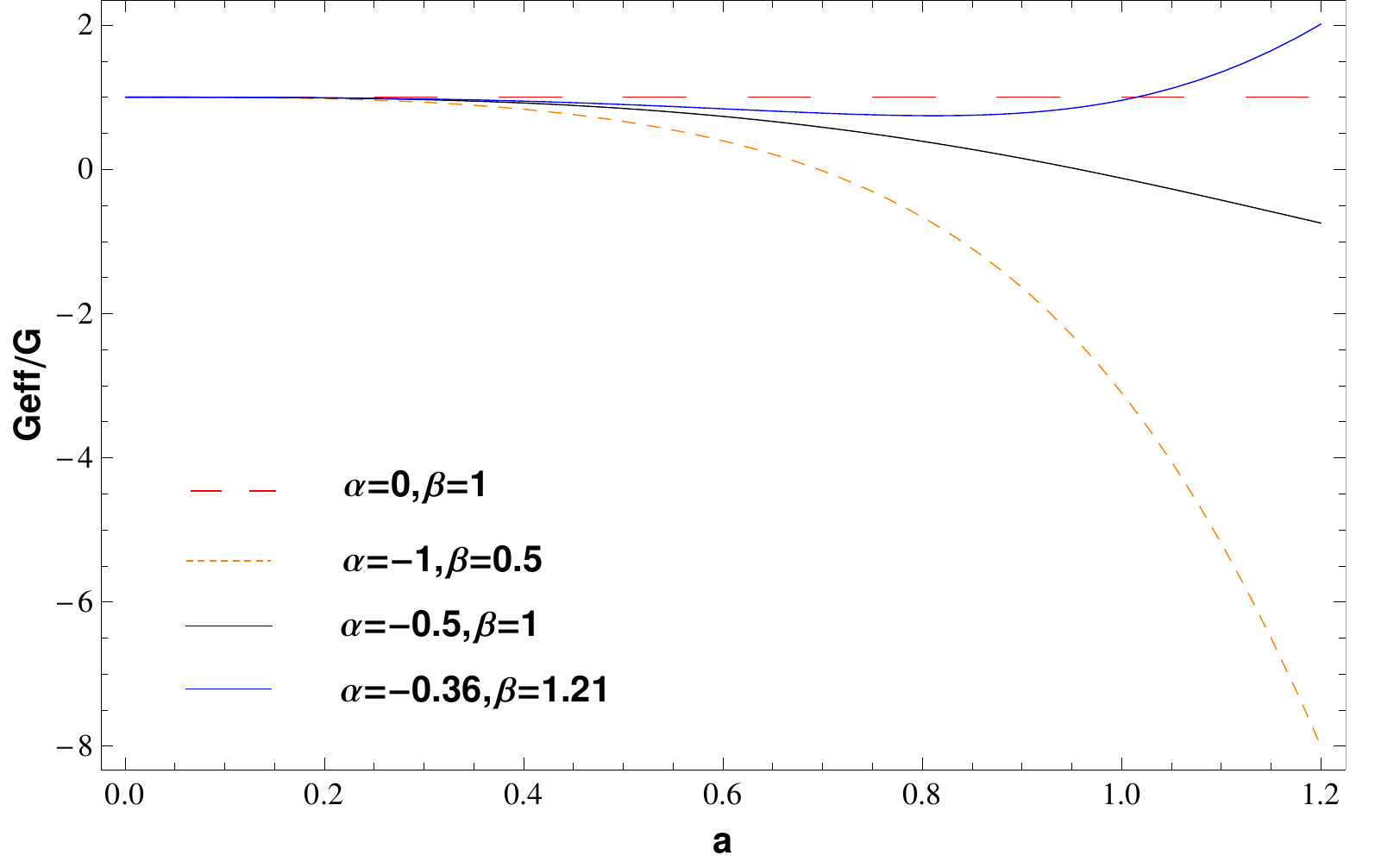}
\includegraphics[scale=0.46]{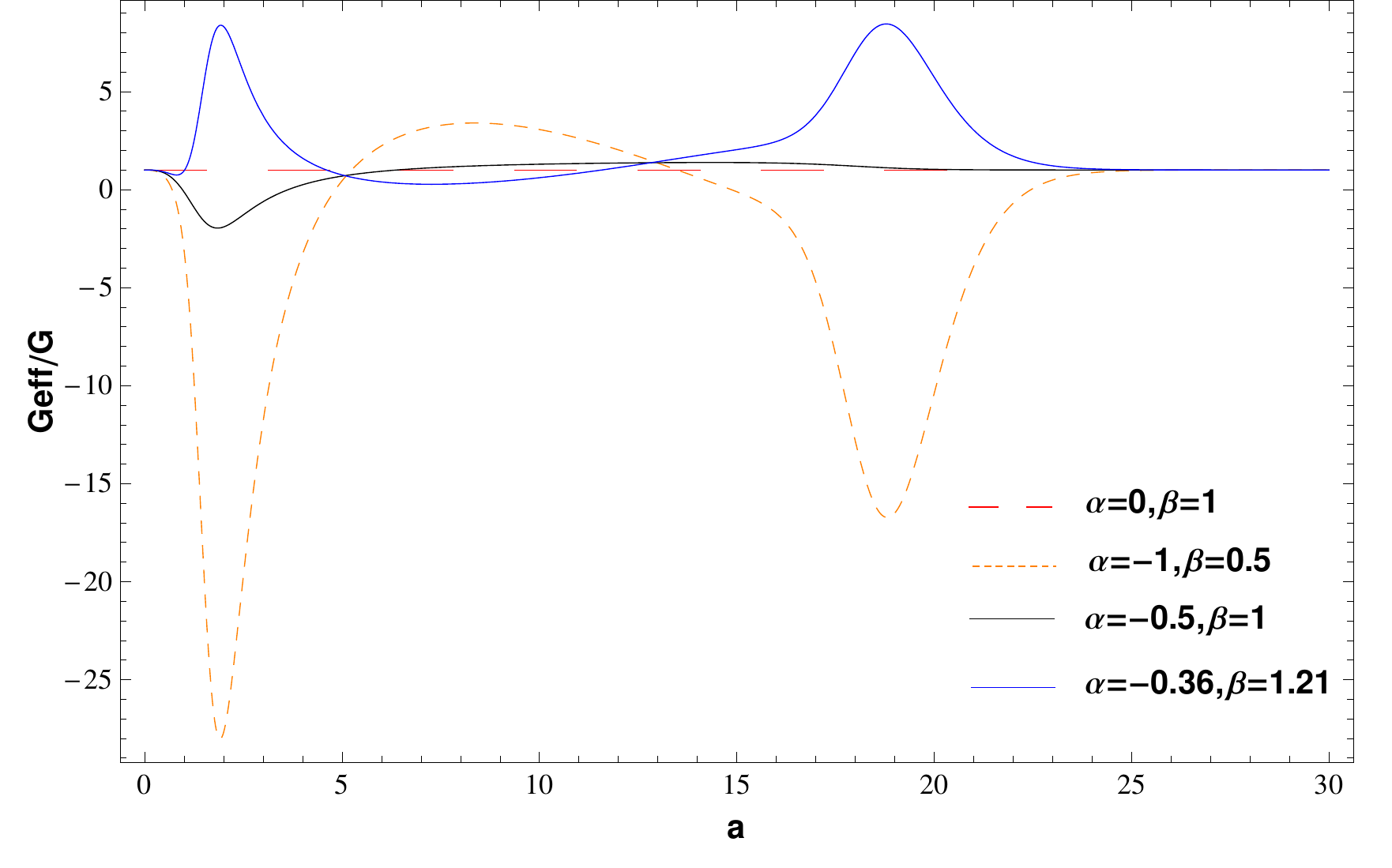}
\caption{The ratio $G_{eff}/G$ as function of the scale factor for different
values of $\alpha$ and $\beta$. The solid blue lines correspond to the best fit. Left panel: evolution  from the past until $a=1.2$. Right panel: future evolution.
Recall that the dark-energy density becomes negative for $a>\sqrt{2/3}\sigma \approx 6$, but is exponentially suppressed. The ratio $G_{eff}/G$ approaches the asymptotic limit $G_{eff}/G =1$ after passing through a second maximum at
$a \approx 19$.}
\label{figGeff}
\end{figure}

\begin{figure}[h]
\includegraphics[scale=0.60]{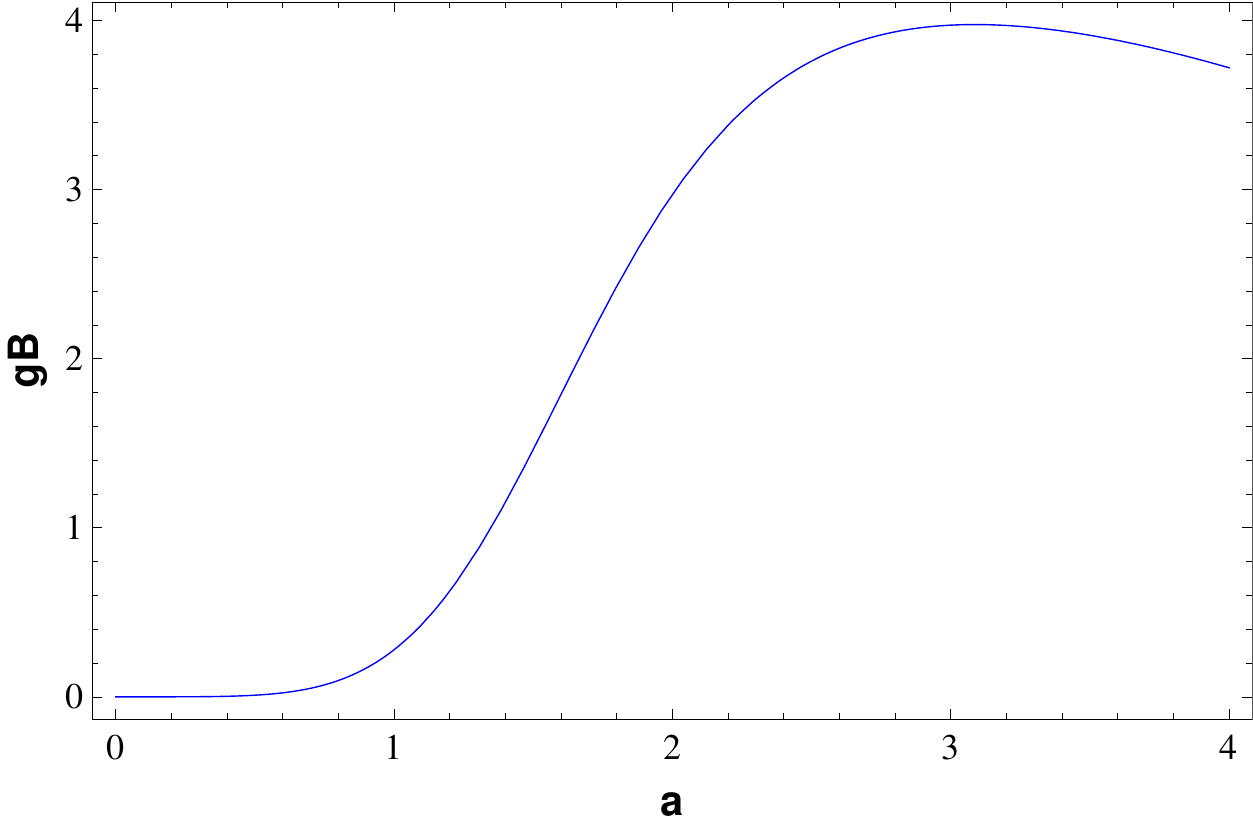}
\includegraphics[scale=0.62]{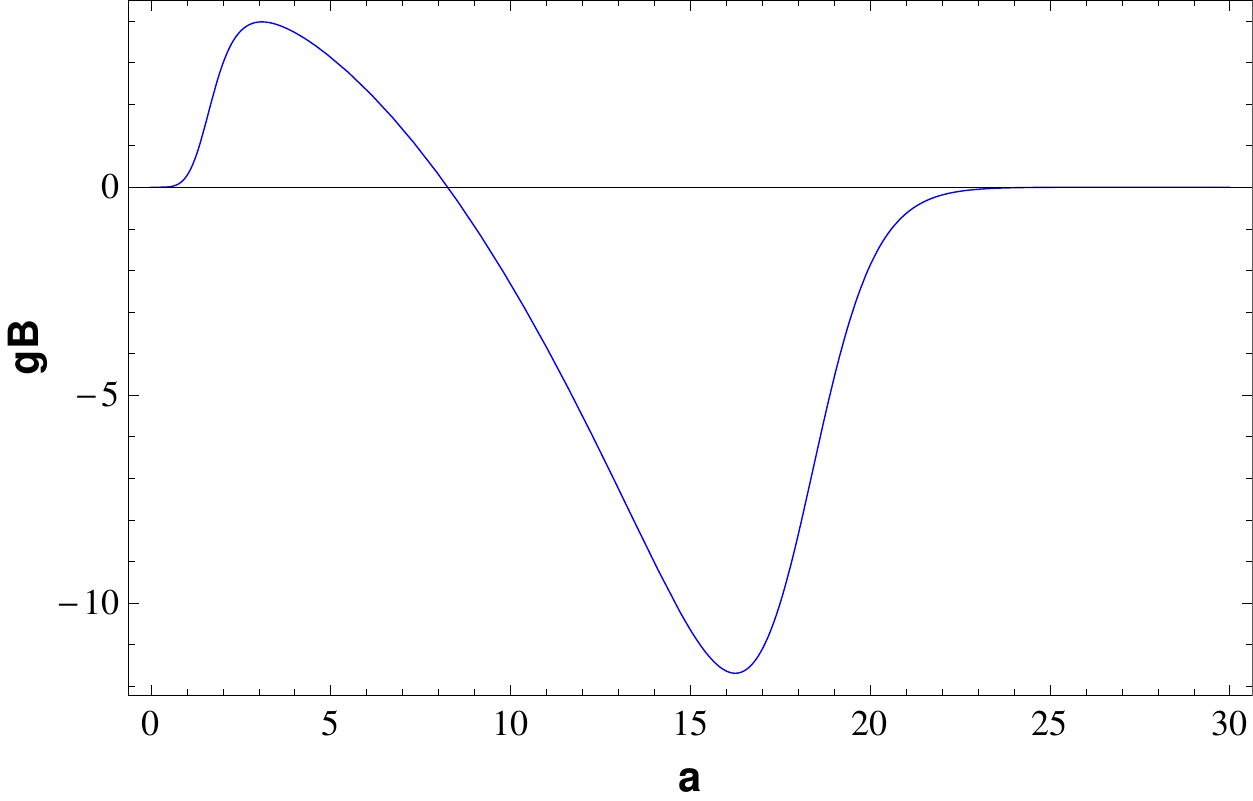}
\caption{The quantity  $gB$ as function of the scale factor for the best-fit
values of $\alpha$ and $\beta$. Left panel: evolution from the past until $a=4$. Right panel: future evolution.
Recall that the  dark-energy density becomes negative for $a>\sqrt{2/3}\sigma \approx 6$, but is exponentially suppressed. The quantity $gB$ approaches the asymptotic limit $gB =0$ after passing through a minimum at
$a \approx 17$.}
\label{figgB}
\end{figure}

\begin{figure}[h]
\includegraphics[scale=0.80]{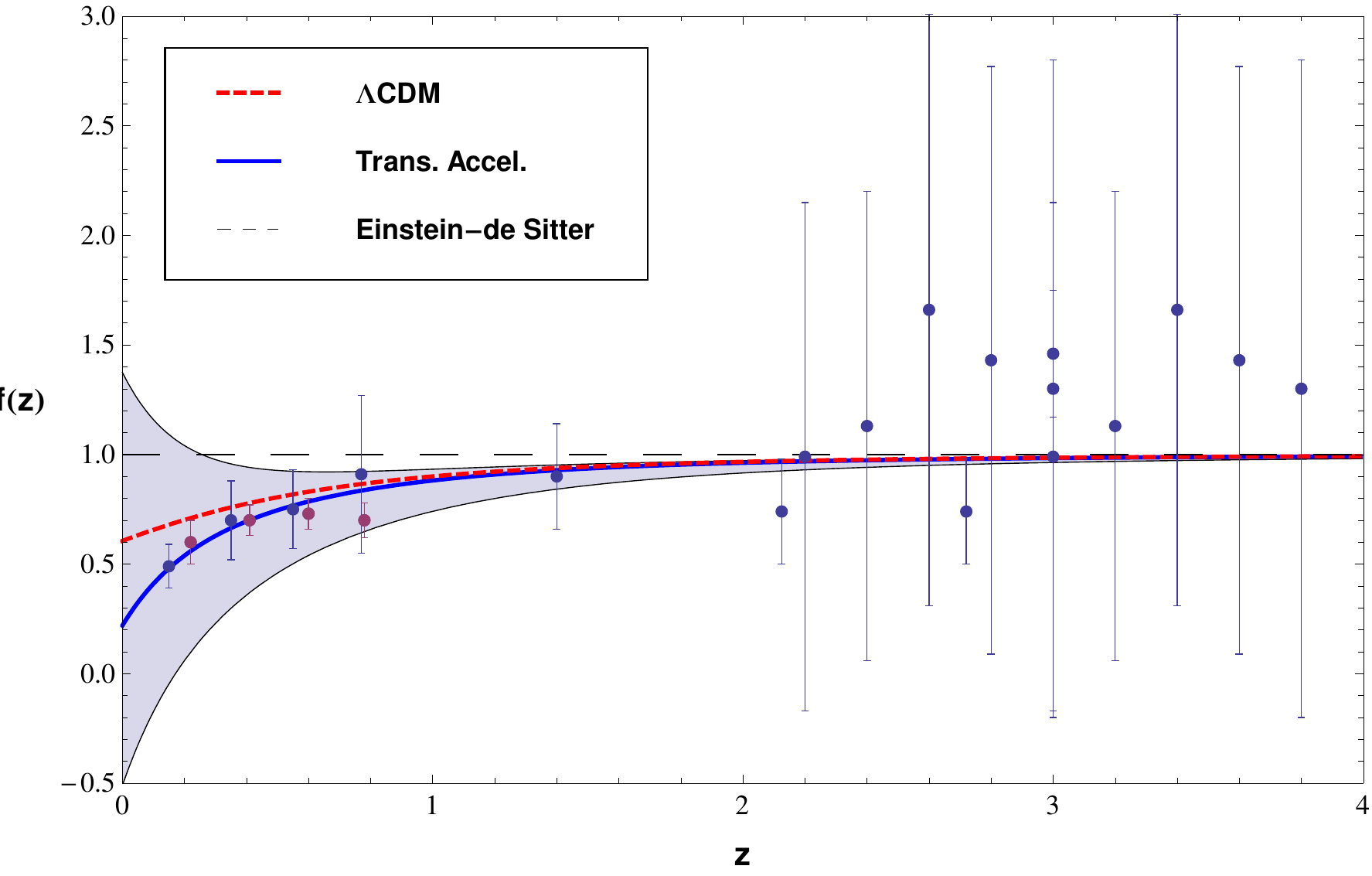}
\caption{Dependence of the growth rate  $f(z)$ on the redshift $z$.
The data are taken from \cite{gong} (blue data points) and \cite{blake} (red data points). The shaded region denotes the $1\sigma$ level, indicating a large dispersion.}
\label{figgrowthf}
\end{figure}

In general, the effective gravitational constant $G_{eff}$ differs from $G$ due to the interaction terms.
The first correction term (parameter $\alpha$) in (\ref{Geff}) describes the direct coupling to the dark-energy fluctuations, the second term encodes the modifications due to the perturbed interaction quantity $\hat{Q}$. For $\beta =1$ the second contribution vanishes: an interaction which  is only operative in the background does not modify the effective gravitational ``constant". In such a case there is an influence of the interaction on $f$ only through the quantity $A(a)$ in the coefficient $U(a)$ (cf.~Eqs.~(\ref{U}) and (\ref{A})).
The effective gravitational ``constant" $G_{eff}$ approaches $G$ in the early matter-dominated phase $a\ll 1$ and also for $a\gg 1$, where the matter dominates again.
The behavior of $G_{eff}$ is shown in Fig.~\ref{figGeff}. The left panel shows the evolution from small values of $a$ in the past until $a=1.2$. The future behavior is depicted in the right panel. Recall that for $a>\sqrt{2/3}\sigma \approx 6$ the dark energy density becomes negative but it is exponentially suppressed. After passing through a second maximum, in the far future $a \gtrsim 22$ one has  $G_{eff} = G$ again.
Fig.~\ref{figgB} shows the corresponding behavior for the best-fit values of the quantity $gB$ which modifies the friction term in the perturbation equation.
In Fig.~\ref{figgrowthf} the growth rate is contrasted with the observations summarized in \cite{gong} as well as with those of \cite{blake}
and with the $\Lambda$CDM model. Around the present epoch ($a\approx 1$) the deviation from the Einstein-de Sitter value is  larger  than that for the $\Lambda$CDM model. This corresponds to the slower growth of $\delta_{m}(a)$ for values of the order of $a\approx 1$ in Fig.~\ref{figdelta1}. For larger values of $a$, however, $\delta_{m}(a)$  continues to grow
while one has  $\delta_{m}(a) = $ const for the $\Lambda$CDM model.

\section{Relativistic perturbation theory}
\label{rel}

\subsection{General relations}

We assume that the cosmic medium as a whole can be described by the energy momentum
tensor of a perfect fluid (neglecting anisotropic stresses and energy fluxes in the rest frame),
\begin{equation}
T_{ik} = \rho u_{i}u_{k} + p h_{ik}\ , \qquad T_{\ ;k}^{ik} = 0,\
\label{T}
\end{equation}
where $h _{ik}=g_{ik} + u_{i}u_{k}$ and $g_{ik}u^{i}u^{k} = -1$. The quantity $u^{i}$ denotes the total four-velocity of the cosmic substratum. Latin indices run from $0$ to $3$.

We assume a split of $T_{ik}$ into a matter component (subindex m) and a dark energy component (subindex x),
\begin{equation}\label{Ttot}
T^{ik} = T_{m}^{ik} + T_{x}^{ik}\ .
\end{equation}
Both these contributions are assumed to have a perfect-fluid structure as well, i.e.,
\begin{equation}\label{TA}
T_{m}^{ik} = \rho_{m} u_m^{i} u^{k}_{m} + p_{m} h_{m}^{ik} \
,\qquad\ h_{m}^{ik} = g^{ik} + u_m^{i} u^{k}_{m} \ .
\end{equation}
and
\begin{equation}\label{TA}
T_{x}^{ik} = \rho_{x} u_A^{i} u^{k}_{x} + p_{x} h_{x}^{ik} \
,\qquad\ h_{x}^{ik} = g^{ik} + u_x^{i} u^{k}_{x} \ .
\end{equation}
Next, we admit an interaction between both components according to
\begin{equation}\label{Q}
T_{m\ ;k}^{ik} = Q^{i},\qquad T_{x\ ;k}^{ik} = - Q^{i}\ .
\end{equation}
Then, the separate energy-balance equations are (cf. \cite{saulo})
\begin{equation}
-u_{mi}T^{ik}_{m\ ;k} = \rho_{m,a}u_{m}^{a} +  \Theta_{m} \left(\rho_{m} + p_{m}\right) = -u_{ma}Q^{a}\ ,
\qquad (g_{ik}u_{m}^{i}u_{m}^{k} = -1)
\label{eb1}
\end{equation}
and
\begin{equation}
-u_{xi}T^{ik}_{x\ ;k} = \rho_{x,a}u_{x}^{a} +  \Theta_{x} \left(\rho_{x} + p_{x}\right) = u_{xa}Q^{a}\ ,
\quad\qquad (g_{ik}u_{x}^{i}u_{x}^{k} = -1)\ .
\label{eb2}
\end{equation}
Each component has its own four-velocity. The quantities $\Theta_{m}$ and $\Theta_{x}$ are defined as $\Theta_{m} = u^{a}_{m;a}$ and $\Theta_{x} = u^{a}_{x;a}$, respectively . For the homogeneous and isotropic background we assume $u_{m}^{a} = u_{x}^{a} = u^{a}$. Likewise, we have the momentum balances
\begin{equation}
h_{mi}^{a}T^{ik}_{m\ ;k} = \left(\rho_{m} + p_{m}\right)a_{m}^{a} + p_{m,i}h_{m}^{ai} = h_{m i}^{a} Q^{i}\
\label{mb1}
\end{equation}
and
\begin{equation}
h_{xi}^{a}T^{ik}_{x\ ;k} = \left(\rho_{x} + p_{x}\right)a_{x}^{a} + p_{x,i}h_{x}^{ai} = - h_{x i}^{a} Q^{i},\
\label{mb2}
\end{equation}
where $a_{m}^{a} \equiv u_{m ;b}^{a}u_{m}^{b}$ and $a_{x}^{a} \equiv u_{x ;b}^{a}u_{x}^{b}$.
The source term $Q^{i}$ is split into parts proportional and perpendicular to the total four-velocity according to
\begin{equation}
Q^{i} = u^{i}Q + \bar{Q}^{i}\ ,
\label{Qdec}
\end{equation}
where $Q = - u_{i}Q^{i}$ and $\bar{Q}^{i} = h^{i}_{a}Q^{a}$, with $u_{i}\bar{Q}^{i} = 0$.

\subsection{The case $p_{x} = - \rho_{x}$}

The contribution $T_{x}^{ik}$ is supposed to describe some form of dark energy. In the simple case of an equation of state $p_{x} = - \rho_{x}$, where $\rho_{x}$ is not necessarily constant, we have
\begin{equation}
T_{x}^{ik} = - \rho_{x}g^{ik}\ .
\label{Tx}
\end{equation}
In the background, the balances (\ref{eb1}) and (\ref{eb2}) take the forms
\begin{equation}
\dot{\rho}_{m} + 3H\rho_{m} = Q^{0}
\label{eb1+}
\end{equation}
and
\begin{equation}
\dot{\rho}_{x}  = - Q^{0}\ ,
\label{eb2+}
\end{equation}
respectively.
Denoting first-order perturbations again by a hat symbol and recalling that for the background $u_{m}^{a} = u_{x}^{a} = u^{a}$ is valid, the perturbed time components of the four-velocities are
\begin{equation}
\hat{u}_{0} = \hat{u}^{0} = \hat{u}_{m}^{0} =\hat{u}_{x}^{0}  = \frac{1}{2}\hat{g}_{00}\ .
\label{u0}
\end{equation}
According to the perfect-fluid structure of both the total energy-momentum tensor (\ref{T}) and the energy-momentum tensors of the components in (\ref{TA}), and with $u_{m}^{a} = u_{x}^{a} = u^{a}$ in the background, we have first-order energy-density perturbations
$\hat{\rho} = \hat{\rho}_{m} + \hat{\rho}_{x}$, pressure perturbations $\hat{p} = \hat{p}_{m} + \hat{p}_{x} = \hat{p}_{x}$
and
\begin{equation}
\hat{T}^{0}_{\alpha} = \hat{T}^{0}_{m\alpha} + \hat{T}^{0}_{x\alpha}\quad\Rightarrow\quad
\left(\rho + p\right)\hat{u}_{\alpha} = \rho_{m}\hat{u}_{m\alpha} + \left(\rho_{x} + p_{x}\right)\hat{u}_{x\alpha}
\ .
\label{T0al}
\end{equation}
Greek indices run from $1$ to $3$. For $p_{x} = - \rho_{x}$ it follows
\begin{equation}
p_{x} = - \rho_{x} \quad\Rightarrow\quad \rho + p = \rho_{m} \quad\Rightarrow\quad\hat{u}_{m\alpha} = \hat{u}_{\alpha}\ .
\label{ual}
\end{equation}
Since the component $m$ is supposed to describe matter, it is clear from (\ref{T0al}) that the perturbed matter velocity $\hat{u}_{m\alpha}$ coincides with the total velocity perturbation $\hat{u}_{\alpha}$.
With $u^{n}_{m} = u^{n}$ up to first order, the energy balance in (\ref{eb1}) (correct up to first order) can be written as
\begin{equation}
\rho_{m,a}u^{a} = -  \Theta \rho_{m}  -u_{a}Q^{a}\ .
\label{eb1u}
\end{equation}
On the other hand, the total energy balance is
\begin{equation}
\rho_{,a}u^{a} = - \Theta \left(\rho + p\right)
\ .
\label{eb}
\end{equation}
For the difference it follows that
\begin{equation}
\dot{\rho} - \dot{\rho}_{m} \equiv \left(\rho - \rho_{m} \right)_{,a}u^{a}
= u_{a}Q^{a}
\ .
\label{diffeb}
\end{equation}
Since, at least up to linear order, $\rho - \rho_{m} = \rho_{x}$, equation (\ref{diffeb}) is equivalent (up to the first order) to
\begin{equation}
\dot{\rho}_{x} \equiv \rho_{x,a}u^{a} = u_{a}Q^{a}
\ .
\label{drx}
\end{equation}
In zeroth order we recover (\ref{eb2+}).
The first-order equation is (cf. (\ref{u0}))
\begin{equation}
\dot{\hat{\rho}}_{x} + \dot{\rho}_{x}\hat{u}^{0} = \widehat{\left(u_{a}Q^{a}\right)}
\ .
\label{diffepert}
\end{equation}
Notice that (\ref{diffepert}) results from a combination of the total energy conservation and the matter energy balance. It has to be consistent with the dark energy balance  (\ref{eb2}). At first order, the latter becomes
\begin{equation}
\dot{\hat{\rho}}_{x} + \dot{\rho}_{x}\hat{u}^{0} = \widehat{\left(u_{xa}Q^{a}\right)}
\ .
\label{eb2pert}
\end{equation}
This means that
\begin{equation}
\widehat{\left(u_{xa}Q^{a}\right)} = \widehat{\left(u_{a}Q^{a}\right)}
\ ,
\label{u2u}
\end{equation}
i.e., the projections of $Q^{a}$ along $u_{xa}$ and along $u_{a}$ coincide. Explicitly,
\begin{equation}
\widehat{\left(u_{a}Q^{a}\right)} = \widehat{\left(u_{a}u^{a}Q\right)} = - \hat{Q}
\ .
\label{uQpert}
\end{equation}
In a next step we consider the momentum balances. The total momentum conservation is described by
\begin{equation}
h_{i}^{a}T^{ik}_{\ ;k} = \left(\rho_{m} + \rho_{x} + p_{x}\right)a^{a} + h_{}^{ai} p_{x, i} = 0\ .
\label{}
\end{equation}
With $p_{x} = - \rho_{x}$  we have
\begin{equation}
h_{i}^{a}T^{ik}_{\ ;k} = \rho_{m}a^{a} + h_{}^{ai} p_{x, i} = 0\ .
\label{mb}
\end{equation}
Using $u^{n}_{m} = u^{n}$ again, the momentum balance (\ref{mb1}) for the matter component  becomes
\begin{equation}
h_{i}^{a}T^{ik}_{m\ ;k} = \rho_{m}a^{a} = h^{ai}Q_{i} = \bar{Q}^{a} = - h_{}^{ai} p_{x, i}\ .
\label{mb1+}
\end{equation}
Notice that we have only used the total momentum conservation and the matter momentum balance.
The momentum balance (\ref{mb2}) of the dark energy degenerates for the case $p_{X} = - \rho_{X}$. It does not describe any dynamics.

Again we introduce the fractional perturbation $\delta_{m} \equiv \frac{\hat{\rho}_{m}}{\rho_{m}}$ in terms of which
the first-order energy balance takes the form
\begin{equation}
\dot{\delta}_{m} + \frac{Q}{\rho_{m}}\delta_{m}  - \phi \left(- 3H + \frac{Q}{\rho_{m}}\right) + \hat{\Theta}
 = \frac{\hat{Q}}{\rho_{m}}\ .
\label{BE1}
\end{equation}
With
\begin{equation}\label{hatu}
\hat{a}_{\alpha}   = \hat{u}_{\alpha ,0} - \frac{1}{2}\hat{g}_{00,\alpha}
= \hat{u}_{\alpha ,0} + \phi_{,\alpha}
\end{equation}
and $u_{\alpha} = v_{,\alpha}$ one finds from the momentum balance (cf. \cite{vdf})
\begin{equation}
\dot{v}_{,\alpha} + \phi_{,\alpha}  = - \frac{1}{\rho_{m}}\left[\hat{p}_{x,\alpha} + \dot{p}_{x}v_{,\alpha}\right] \quad \Rightarrow\quad \dot{v} + \phi = - \frac{1}{\rho_{m}}\left[\hat{p}_{x} + \dot{p}_{x}v\right]
\ .
\label{BM1}
\end{equation}
At this stage the relation to the previous Newtonian treatment becomes evident.
Neglecting the perturbations on the right-hand side of the momentum balance in (\ref{BM1}) and replacing
$v \rightarrow a v$ we recover the result (\ref{euler}) of the Newtonian analysis.
Neglecting the terms multiplied by $\phi$ in (\ref{BE1}) and identifying $\hat{\Theta}$ accordingly, the non-relativistic relation (\ref{dotdelta}) is reproduced.

The dynamics of the expansion scalar is determined by the
Raychaudhuri equation which in our case takes the form
\begin{equation}
\dot{\Theta} + \frac{1}{3}\Theta^{2} - a^{a}_{;a} + 4\pi G \left(\rho + 3
p\right) = 0\ .\label{Ray}
\end{equation}
In the background, $\dot{H} = -4\pi G\left(\rho + p\right) = - 4\pi G \rho_{m}$ is valid.
The term $a^{m}_{;m}$ in the Raychaudhuri equation becomes at first order
\begin{equation}\label{dotumm}
a^{m}_{;m} = -\frac{1}{a^{2}\rho_{m}}\left(\Delta \hat{p}_{x} + \dot{p}_{x}\Delta v\right)\ ,
\end{equation}
where $\Delta$ is the three-dimensional Laplacian.

For the perturbed time derivative of the expansion scalar we have
\begin{equation}\label{hdtheta}
\hat{\dot{\Theta}} = \dot{\hat{\Theta}} + \dot{\Theta}\hat{u}^{0} = \dot{\hat{\Theta}} - \dot{\Theta}\phi\ .
\end{equation}
Consequently, the first two terms of the perturbed Raychaudhuri equation are
\begin{equation}\label{hatdt+}
\left[\dot{\Theta} + \frac{1}{3}\Theta^{2}\right]^{\mathbf{\hat{}}} = \dot{\hat{\Theta}} - \dot{\Theta}\phi + \frac{2}{3}\Theta\hat{\Theta}\ .
\end{equation}
For the derivative of the expansion scalar $\dot{\Theta} = - 12\pi G\rho_{m}$ is valid.
For the perturbations of the term  $4\pi G \left[\rho + 3p\right]$ we find
\begin{equation}\label{}
4\pi G \left[\rho + 3p\right]^{\mathbf{\hat{}}} = 4\pi G \left(\rho_{m}\delta_{m} + \rho_{x}\delta_{x} + 3 \hat{p}_{x}\right)\ .
\end{equation}

\subsection{The Gauge-invariant perturbation equation}

It is convenient now to introduce gauge-invariant quantities to describe the perturbation dynamics by
\begin{equation}\label{defdc}
\delta_{m}^{c} = \delta_{m} + \frac{\dot{\rho}_{m}}{\rho_{m}}v \ , \qquad \delta_{x}^{c} = \delta_{x} + \frac{\dot{\rho}_{x}}{\rho_{x}}v\ , \qquad \hat{p}_{x}^{c} = \hat{p}_{x} + \dot{p}_{x}v
\end{equation}
as well as
\begin{equation}\label{Qc}
\hat{\Theta}^{c} = \hat{\Theta} + \dot{\Theta}v\ , \qquad \mathrm{and} \qquad \hat{Q}^{c} = \hat{Q} + \dot{Q}v\ .
\end{equation}
The superscript $c$ stands for comoving. All the symbols have their physical meaning on comoving hypersurfaces $v=0$.
From  (\ref{BM1}) with $\hat{p}^{c}_{x} = c_{s}^{2}\hat{\rho}_{x}^{c}$, where $c_{s}$ is the sound speed in the rest-frame $v=0$, it  follows that
\begin{equation}\label{dotvx}
\dot{v} + \phi = - c_{s}^{2}\frac{\rho_{x}}{\rho_{m}}\delta_{x}^{c}\ .
\end{equation}
Equation (\ref{BE1}) takes the form
\begin{equation}\label{dmc}
\dot{\delta}_{m}^{c} + \frac{Q}{\rho_{m}}\delta_{m}^{c} + \hat{\Theta}^{c} + c_{s}^{2} \frac{\dot{\rho}_{m}}{\rho_{m}}\frac{\rho_{x}}{\rho_{m}}\delta_{x}^{c} = \frac{1}{\rho_{m}}\hat{Q}^{c}\ .
\end{equation}
Since
\begin{equation}\label{hQdr}
\hat{Q} = - \hat{\dot{\rho}}_{x}\ ,
\end{equation}
we may rewrite $Q^{c}$ as
\begin{equation}\label{hQdr}
\hat{Q}^{c} = - \rho_{x}\dot{\delta}_{x}^{c} - \dot{\rho}_{x}\left[1 + c_{s}^{2}\frac{\rho_{x}}{\rho_{m}}\right]\delta_{x}^{c}\ .
\end{equation}
The gauge-invariant perturbation of the interaction quantity $Q$ is determined by the dark-energy density perturbations and their first derivative.
It follows that
\begin{equation}\label{dmc2}
\dot{\delta}_{m}^{c} + \hat{\Theta}^{c} + \frac{Q}{\rho_{m}}\delta_{m}^{c}
- \left(3H c_{s}^{2}\frac{\rho_{x}}{\rho_{m}} + \frac{Q}{\rho_{m}}\right)\delta_{x}^{c} = - \frac{\rho_{x}}{\rho_{m}}\dot{\delta}_{x}^{c}
\ .
\end{equation}
In terms of the gauge-invariant quantities, the first-order Raychaudhuri equation becomes
\begin{equation}\label{rayc}
\dot{\hat{\Theta}}^{c} = - \frac{2}{3}\Theta\hat{\Theta}^{c} - \frac{1}{a^{2}\rho_{m}}\Delta\hat{p}_{x}^{c}
- 4\pi G\left[\rho_{m}\delta_{m}^{c} + \rho_{x}\delta_{x}^{c}\right]\ .
\end{equation}
In a next step we differentiate (\ref{dmc2}) and use
\begin{equation}\label{dotratio}
\left(\frac{\rho_{x}}{\rho_{m}}\right)^{\displaystyle\cdot} = \frac{\rho_{x}}{\rho_{m}}
\left[3H - \frac{Q}{\rho_{m}}\frac{\rho}{\rho_{x}}\right]
\end{equation}
and
\begin{equation}\label{dH2}
\dot{H} = - 4\pi G\rho_{m} = - \frac{3}{2}H^{2}\frac{\rho_{m}}{\rho}\
\end{equation}
as well as (\ref{rayc}). The result is
\begin{eqnarray}
\ddot{\delta}_{m}^{c} &+& \left(2H + \frac{Q}{\rho_{m}}\right)\dot{\delta}_{m}^{c}
+ \left[\left(\frac{Q}{\rho_{m}}\right)^{\displaystyle\cdot} + 2H\frac{Q}{\rho_{m}}\right]\delta_{m}^{c}
- 4\pi G\rho_{m}\delta_{m}^{c} \nonumber\\
 &=&  - \frac{\rho_{x}}{\rho_{m}}\ddot{\delta}_{x}^{c}
 - \left[\left(5 - 3c_{s}^{2}\right)H\frac{\rho_{x}}{\rho_{m}} - \left(2 + \frac{\rho_{x}}{\rho_{m}}\right)\frac{Q}{\rho_{m}}\right]\dot{\delta}_{x}^{c}
 + 4\pi G\rho_{x}\delta_{x}^{c} + \frac{1}{a^{2}\rho_{m}}\Delta\hat{p}_{x}^{c}\nonumber\\
 && + \left[\left(\frac{Q}{\rho_{m}}\right)^{\displaystyle\cdot}
 + 3H^{2}c_{s}^{2}\left(5\frac{\rho_{x}}{\rho_{m}} - \frac{3}{2}\frac{\rho_{x}}{\rho}\right)
 + H \frac{Q}{\rho_{m}}
 \left(2 - 3c_{s}^{2}\left(1+ \frac{\rho_{x}}{\rho_{m}}\right)\right)\right]\delta_{x}^{c}\ .
  \label{dddelta}
\end{eqnarray}
Now we change to the scale factor as independent variable according to (\ref{prime}).
Applying
\begin{equation}\label{Hpr}
\frac{H^{\prime}}{H} = - \frac{3}{2a}\frac{\rho_{m}}{\rho}\ ,
\end{equation}
and
\begin{equation}\label{primepr}
\ddot{\delta} = a^{2}H^{2}
\left[\delta^{\prime\prime} + \left(1 - \frac{3}{2}\frac{\rho_{m}}{\rho}\right)\frac{\delta^{\prime}}{a}\right]\ ,
\end{equation}
the previous equation (\ref{dddelta}) can be transformed into
\begin{eqnarray}
\delta_{m}^{c\prime\prime} &+& \left[\frac{3}{2} + \frac{3}{2}\frac{\rho_{x}}{\rho}
+ \frac{Q}{H\rho_{m}}\right]\frac{\delta_{m}^{c\prime}}{a} + \left[-\frac{3}{2} + \frac{3}{2}\frac{\rho_{x}}{\rho} + \frac{1}{H^{2}}\left(\frac{Q}{\rho_{m}}\right)^{\displaystyle\cdot}
+ 2 \frac{Q}{H\rho_{m}}\right]\frac{\delta_{m}^{c}}{a^{2}} \nonumber\\
 &=&  - \frac{\rho_{x}}{\rho_{m}}\delta_{x}^{c\prime\prime}
 - \left[\left(6 - 3 c_{s}^{2}\right) \frac{\rho_{x}}{\rho_{m}} - \frac{3}{2} \frac{\rho_{x}}{\rho}
 - \left(2 + \frac{\rho_{x}}{\rho_{m}}\right)\frac{Q}{H\rho_{m}}\right]\frac{\delta_{x}^{c\prime}}{a}\nonumber\\
&& +\left[\frac{1}{H^{2}}\left(\frac{Q}{\rho_{m}}\right)^{\displaystyle\cdot}
+ \frac{Q}{H\rho_{m}}\left(2 - 3c_{s}^{2}\left(1+\frac{\rho_{x}}{\rho_{m}}\right)\right)
+ 3 c_{s}^{2}\left(5 \frac{\rho_{x}}{\rho_{m}}  -  \frac{3}{2}\frac{\rho_{x}}{\rho}\right) \right.\nonumber\\
&& \qquad + \left.
\frac{3}{2}\frac{\rho_{x}}{\rho} -  c^{2}_{s} \frac{k^{2}}{a^{2}H^{2}}\frac{\rho_{x}}{\rho_{m}}\right]\frac{\delta_{x}^{c}}{a^{2}}\ .\nonumber\\
\end{eqnarray}
Here we have used
\begin{equation}\label{dk}
\frac{\Delta\hat{p}_{x}^{c}}{a^{2}H^{2}\rho_{m}}\quad\rightarrow\quad - c^{2}_{s}\frac{k^{2}}{a^{2}H^{2}}\frac{\rho_{x}}{\rho_{m}}
\delta_{x}^{c}\ .
\end{equation}
We have expressed the perturbations $\hat{Q}^{c}$ of the interaction term in terms of the dark-energy perturbation
$\delta_{x}^{c}$ and its first derivative according to (\ref{hQdr}).
In order to obtain a closed second-order equation for $\delta_{m}^{c}$, we assume a proportionality between the
dark-energy perturbations and the dark-matter perturbation by introducing
\begin{equation}\label{dcm}
\delta_{x}^{c} = \epsilon \delta_{m}^{c}\ .
\end{equation}
The parameter $\epsilon$ quantifies the relative magnitude of the perturbations of the dark energy.
It corresponds to the parameter $\alpha$ of the non-relativistic theory. To avoid misunderstandings, we use a different symbol here.
In many studies dark-energy perturbations are neglected from the outset. However, this is strictly justified only for a cosmological constant. Neglecting these perturbation may lead to unreliable conclusions concerning
the interpretation of observational data \cite{Park-Hwang,ivan}. For the model dealt with in \cite{saulo},
dark-energy perturbations were shown to be negligible on scales that are relevant for structure formation, but may play a role on super-horizon scales.

With (cf. (\ref{dQrm}))
\begin{equation}\label{dotQ}
\frac{1}{H^{2}}\left(\frac{Q}{\rho_{m}}\right)^{\displaystyle\cdot} =
\left[-\frac{3}{2} + A + B\right]\frac{Q}{H\rho_{m}} - 4 \frac{a^{2}}{\sigma^{2}}\frac{g}{1+ g}
\end{equation}
we find
\begin{equation}
\delta_{m}^{c\prime\prime} + F(a)\delta_{m}^{c\prime} + G(a) \delta_{m}^{c} = 0
\ ,
\label{deltaprprrel}
\end{equation}
where
\begin{equation}
F(a) = \frac{1}{a\left(1+\epsilon\frac{\rho_{x}}{\rho_{m}}\right)}
\left\{\frac{3}{2} + \frac{3}{2}\left(1 - \epsilon\right)\frac{\rho_{x}}{\rho} + gB - \epsilon
\left[\left(2 +  \frac{\rho_{x}}{\rho_{m}}\right)gB - \left(6-3c_{s}^{2}\right) \frac{\rho_{x}}{\rho_{m}}\right]
\right\}
\
\label{F(a)}
\end{equation}
and
\begin{eqnarray}
G(a) &=& \frac{1}{a^{2}\left(1+\epsilon\frac{\rho_{x}}{\rho_{m}}\right)}
\left\{-\frac{3}{2} + \frac{3}{2}\left(1-\epsilon\right)\frac{\rho_{x}}{\rho}
+ \left(\frac{1}{2} + A + B\right)gB - 4 \frac{a^{2}}{\sigma^{2}}\frac{g}{1+g}
\right.\nonumber\\
&&\left.
- \epsilon\left[\left(\frac{1}{2} + A + B - 3 c_{s}^{2}
\left(1 + \frac{\rho_{x}}{\rho_{m}}\right)\right)gB -  4 \frac{a^{2}}{\sigma^{2}}\frac{g}{1+g}
\right.\right.\nonumber\\
&&\qquad + \left.\left. 15 c_{s}^{2}\frac{\rho_{x}}{\rho_{m}} - \frac{9}{2} c_{s}^{2}\frac{\rho_{x}}{\rho} - c_{s}^{2}\frac{\rho_{x}}{\rho_{m}}\frac{k^{2}}{a^{2}H^{2}}\right]
\right\}
\ ,
\label{G(a)}
\end{eqnarray}
with
\begin{equation}
\frac{\rho_{x}}{\rho_{m}} = a^{3}\frac{\frac{3}{2}\sigma^{2}\bar{K}\exp{\left(-a^{2}/\sigma^{2}
\right)}\left(1 - \frac{2}{3}\frac{a^{2}}{\sigma^{2}}\right)}{1 +
\bar{K}\left(a^{5}\exp{\left(-a^{2}/\sigma^{2}
\right)} - \frac{3}{2}\sigma^{2}\exp{\left(-1/\sigma^{2}
\right)}\right)}
\
\label{rx/rm}
\end{equation}
and
\begin{equation}
\frac{\rho_{x}}{\rho} = \left[\frac{3}{2}\sigma^{2}\bar{K}\exp{\left(-a^{2}/\sigma^{2}
\right)}\left(1 - \frac{2}{3}\frac{a^{2}}{\sigma^{2}}\right)\right]
\frac{H_{0}^{2}}{H^{2}} = \frac{2}{3}A
\ .
\label{rx/r}
\end{equation}
It is only for $\epsilon\neq 0$ that we have a scale dependence.
The quantities $A(a)$, $\frac{H_{0}^{2}}{H^{2}}$ and $B$, are given by (\ref{A}), (\ref{H0/H}) and (\ref{B}), respectively, with  (cf. (\ref{dotg}))
\begin{equation}\label{dotg+}
g
= \frac{\bar{K}\,a^{5}\,\exp\left(-a^{2}/\sigma^{2}\right)}{\Omega_{m_0} - \bar{K}\,\exp\left(-1/\sigma^{2}\right)}
= \frac{\bar{K}\,a^{5}\,\exp\left(-a^{2}/\sigma^{2}\right)}{1 - \frac{3}{2}\sigma^{2}\bar{K}\,\exp\left(-1/\sigma^{2}\right)} \ ,
\end{equation}
where we have used that $\Omega_{m0} = 1 -  \bar{K}\exp(-1/\sigma^{2})\left(\frac{3}{2}\sigma^{2} -1\right)$.

For $\epsilon = 0$ we recover
\begin{equation}
F(a) = \frac{1}{a}
\left\{\frac{3}{2} + \frac{3}{2}\frac{\rho_{x}}{\rho} + gB
\right\} \qquad\qquad (\epsilon = 0)
\
\label{F(a)0}
\end{equation}
and
\begin{equation}
G(a) = \frac{1}{a^{2}}
\left\{-\frac{3}{2} + \frac{3}{2}\frac{\rho_{x}}{\rho}
+ \left(\frac{1}{2} + A + B\right)gB - 4 \frac{a^{2}}{\sigma^{2}}\frac{g}{1+g}
\right\}\qquad (\epsilon = 0)
\ .
\label{G(a)0}
\end{equation}
These coefficients coincide with the corresponding coefficients of the Newtonian analysis for $\alpha=\beta = 0$. Since $\hat{Q}^{c}$ was related to $\delta_{x}^{c}$ and its first derivative in
(\ref{hQdr}), the limit $\alpha=\beta = 0$ of the Newtonian theory corresponds to $\epsilon = 0$ within the present relativistic analysis.
Recall that $\frac{\rho_{x}}{\rho}= \frac{2}{3}A$.

Our interest is the matter power spectrum, defined by $P_k=\left|\delta_{m,k}\right|^{2}$,
where $\delta_{m,k}$ is the Fourier component of the density contrast $\delta_{m}^{c}$.
In order to choose appropriate initial conditions, we use the circumstance that
at early times, i.e. for small scale factors $a \ll 1$, the equation (\ref{deltaprprrel}) has the asymptotic Einstein-de Sitter form
\begin{equation}\label{Early}
\qquad \qquad \delta_{m}'' + \frac{3}{2a} \,\delta_{m}' - \frac{3}{2a^2}\,\delta_{m} =0 \,,\qquad \qquad  (a \ll 1)\ ,
\end{equation}
which coincides with the corresponding equation of the $\Lambda$CDM model at that period.
This allows us to relate our model to the $\Lambda$CDM model at high redshift.
We shall benefit from the fact that the matter power spectrum for the
$\Lambda$CDM model is well fitted by the BBKS transfer
function \cite{bbks}.
Integrating the $\Lambda$CDM model back from today to a distant past, say $z = 10^{5}$, we
obtain the shape of the transfer function at that moment.
The
spectrum determined in this way is then used as initial
condition for our model. This procedure is similar to that described in
more detail in references \cite{sola,sauloIC}.

\begin{figure}[!h]
\centering
\includegraphics[width=0.80\textwidth]{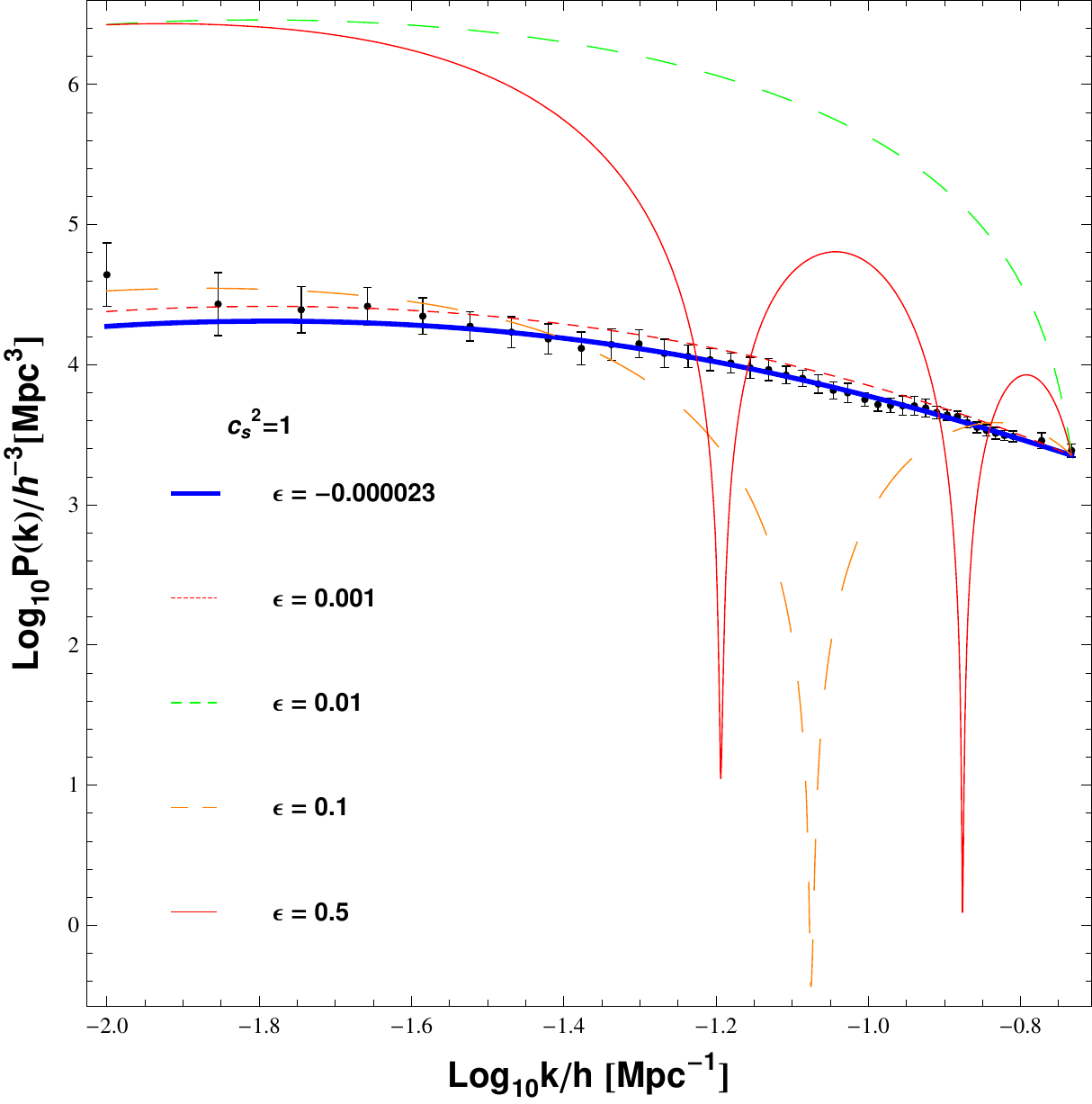}
\caption{Matter power spectrum for $c_{s}^{2} = 1$ and different values of $\epsilon$. The thick solid (blue) curve ($\epsilon = -0.000023$) represents the best overall fit. On large scales, however the curve with $\epsilon = 0.001$ shows the better performance. Larger values of $\epsilon$ result in (non-observed) oscillations. The data are taken from \cite{cole}.}
\label{figespectro1}
\end{figure}

\begin{figure}[h]
\includegraphics[scale=0.7]{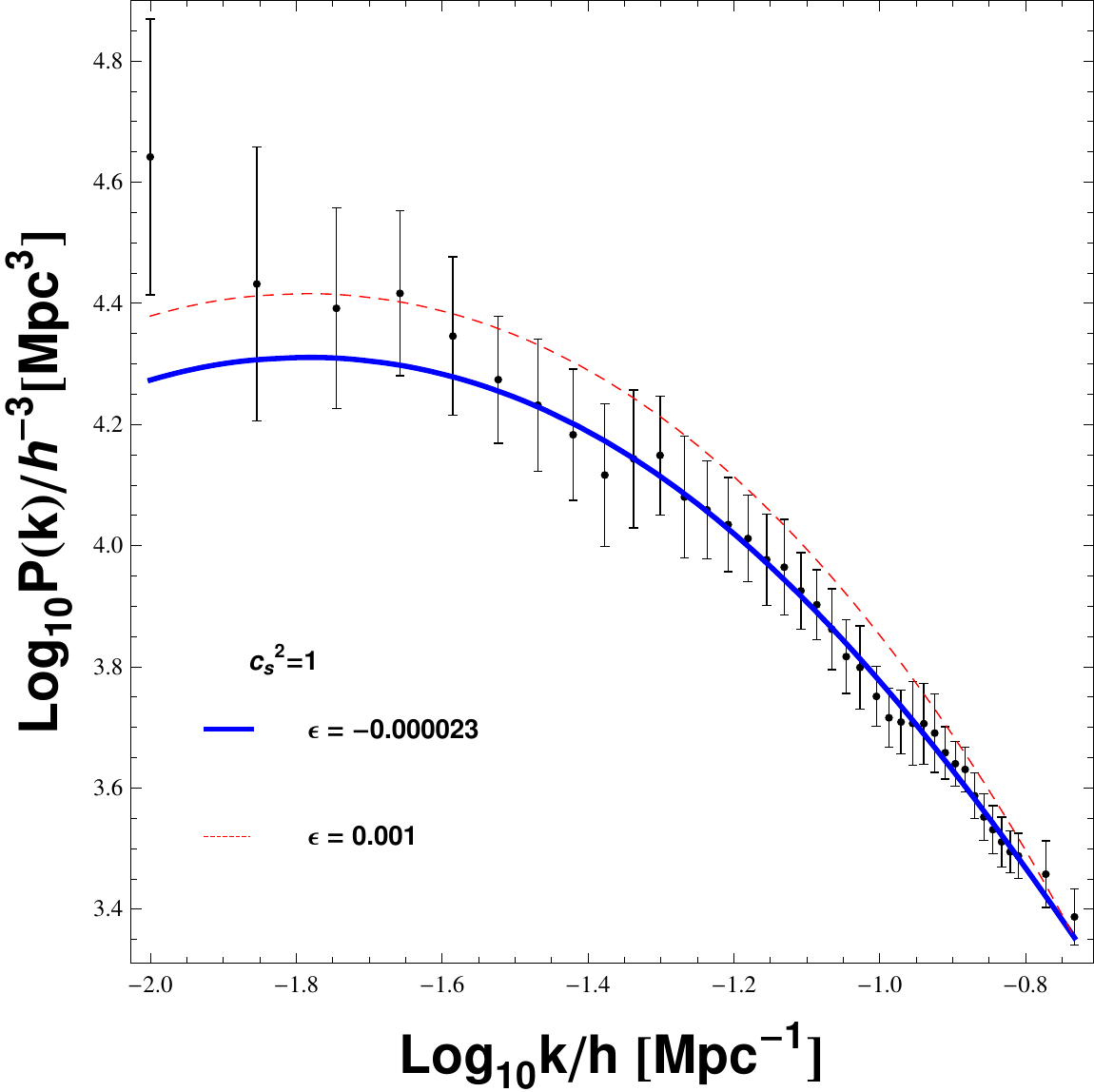}
\includegraphics[scale=0.7]{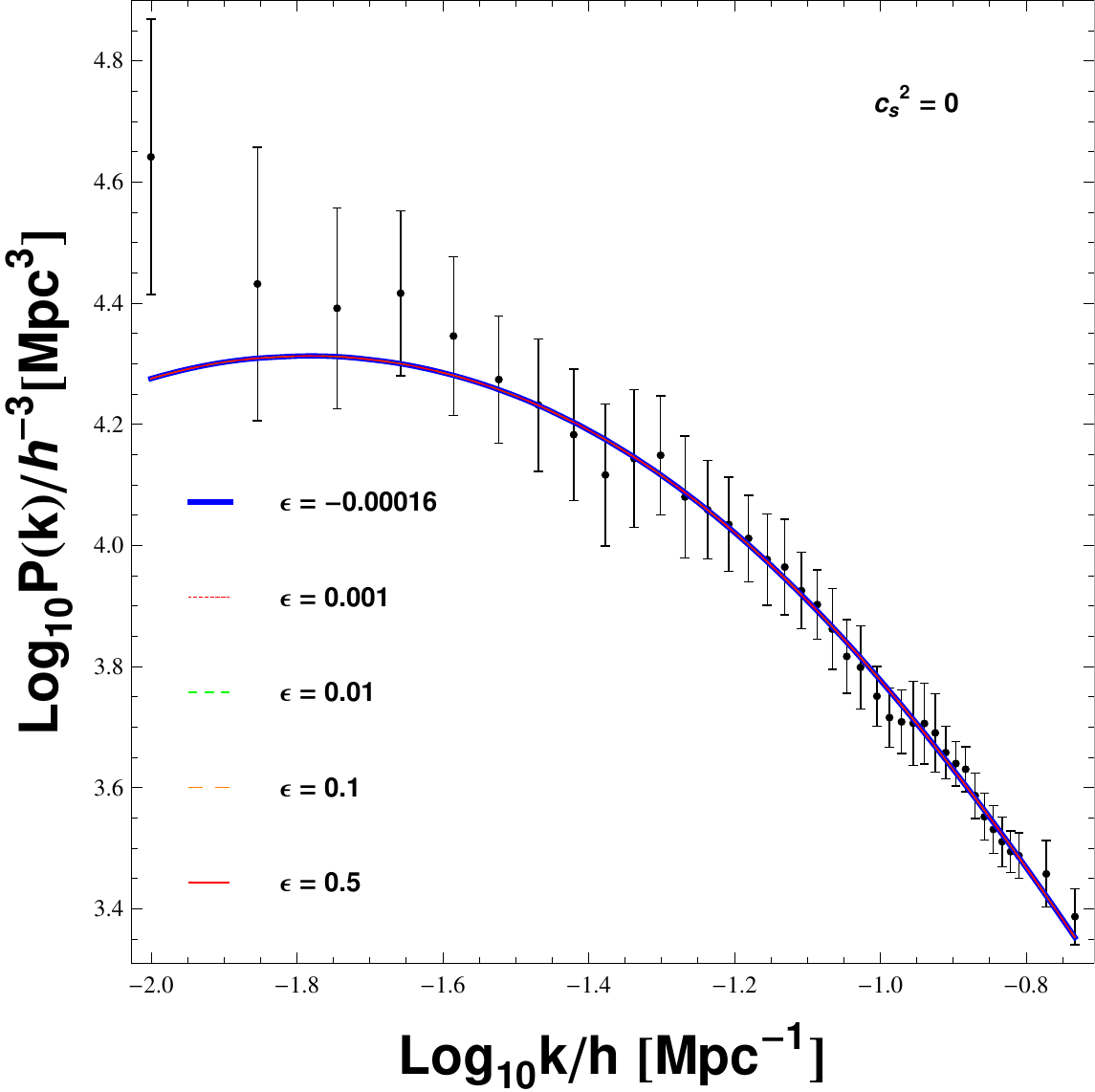}
\caption{Matter power spectrum for $c_{s}^{2} = 1$ (left panel) and $c_{s}^{2} = 0$ (right panel).
While the total best-fit value for $c_{s}^{2} = 1$ is $\epsilon = - 0.000023$, it is obvious that on larger scales the dashed curve with $\epsilon = 0.001$ gives a better description. This corresponds to
the expectation that the dark-energy perturbations are more relevant on the largest scales. For $c_{s}^{2} = 0$ the resulting curves do not depend on $\epsilon$. The data are taken from \cite{cole}.}
\label{Spectrum32}
\end{figure}

In Fig.~\ref{figespectro1} we display the power spectrum, based on the data of the 2dFGRS project, for different values of $\epsilon$ for $c_{s}^{2} = 1$. The scale dependence via the coefficient $G(a)$ in eq.~(\ref{deltaprprrel}) is sensitive to the product $\epsilon c_{s}^{2}$. The thick (blue) curve ($\epsilon = -0.000023$) represents the best overall fit. On large scales, however, the curve with $\epsilon = 0.001$ shows the better performance.
For $c_{s}^{2} = 1$ only a very small factor $\epsilon$ is compatible with the data. Otherwise, there appear non-observed oscillations in the matter power spectrum which resemble
a similar behavior in (generalized) Chaplygin gases \cite{Sandvik,vdf}.
The left panel of Fig.~\ref{Spectrum32} provides an alternative illustration of the matter power spectrum.
The right panel confirms that there is no scale dependence for $c_{s}^{2} = 0$. There is no dependence on $\epsilon$ either. The solid blue line just reproduces the BBKS transfer function.
We verified that also for small non-vanishing values of $c_{s}^{2}$, i.e. $c_{s}^{2} = 0.1$, the resulting best-fit values for $\epsilon$ are of the order $10^{-4}...10^{-3}$. Although a constant $\epsilon$ corresponds to a very rough approximation, these results indicate that fluctuations of the dark-energy component are small indeed on scales that are relevant for galaxy formation. On the other hand, if they were zero exactly,
there would be no scale dependence at all. As the left panel of Fig.~\ref{Spectrum32} shows, even a very small value of $\epsilon$, although considerably larger than the best-fit value, influences the spectrum substantially on larger scales. This demonstrates an increasing role of the dark-energy perturbations with increasing scale. The point here is that the much larger number of data for the smallest scales has more weight in the statistical analysis than the fewer data on larger scales. Obviously, the same constant value of (an almost vanishing) $\epsilon$ which gives a correct description on the smallest scales is not adequate on the larger scales of the sample (left panel of Fig.~\ref{Spectrum32}).
Conversely, a good performance on large scales is not compatible with the observations on intermediate and small scales. Consequently, for a more advanced analysis, a scale-dependent $\epsilon$ should be used.

\subsection{Non-adiabatic perturbations}

\begin{figure}[h]
\includegraphics[scale=0.65]{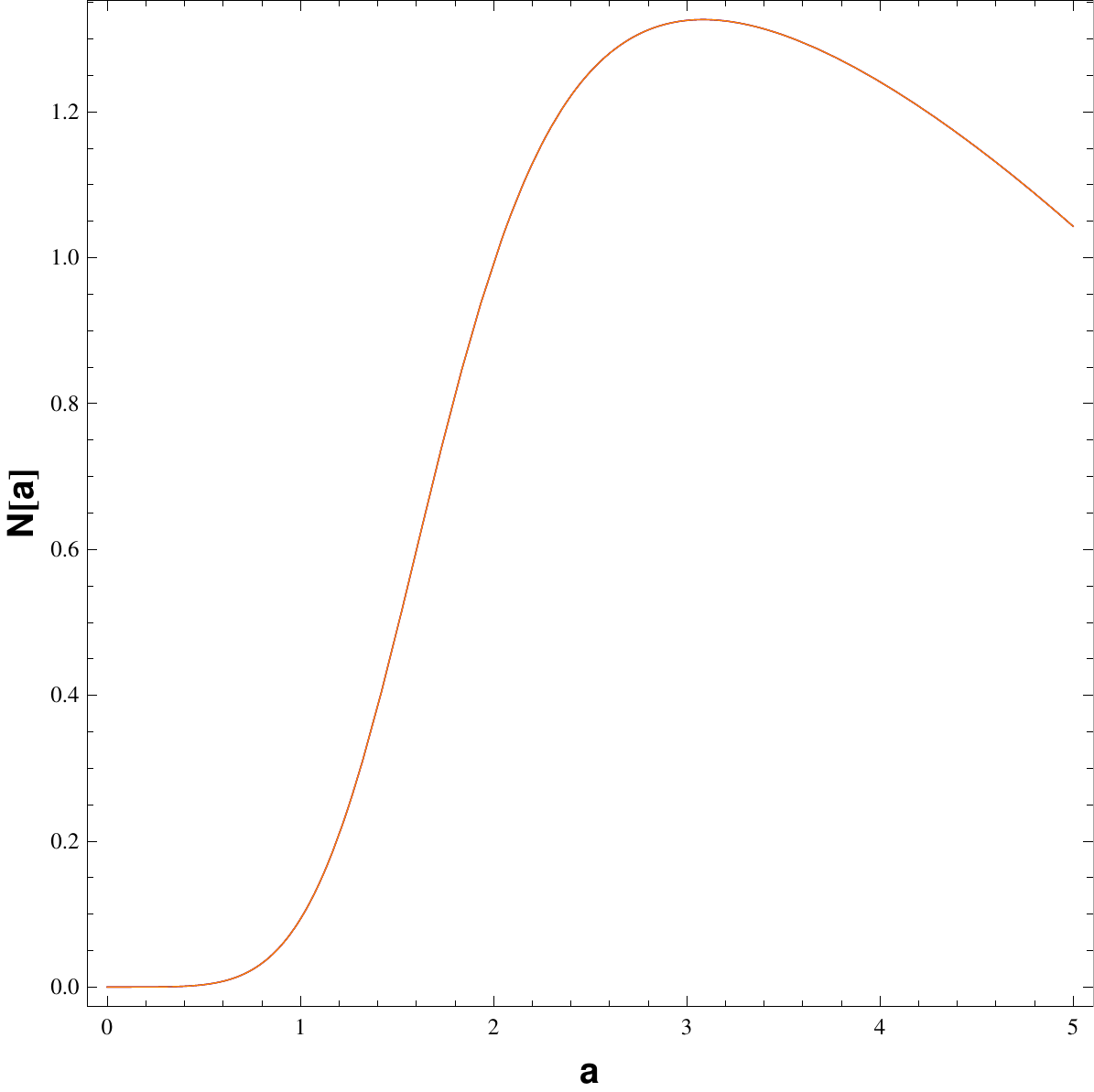}
\includegraphics[scale=0.75]{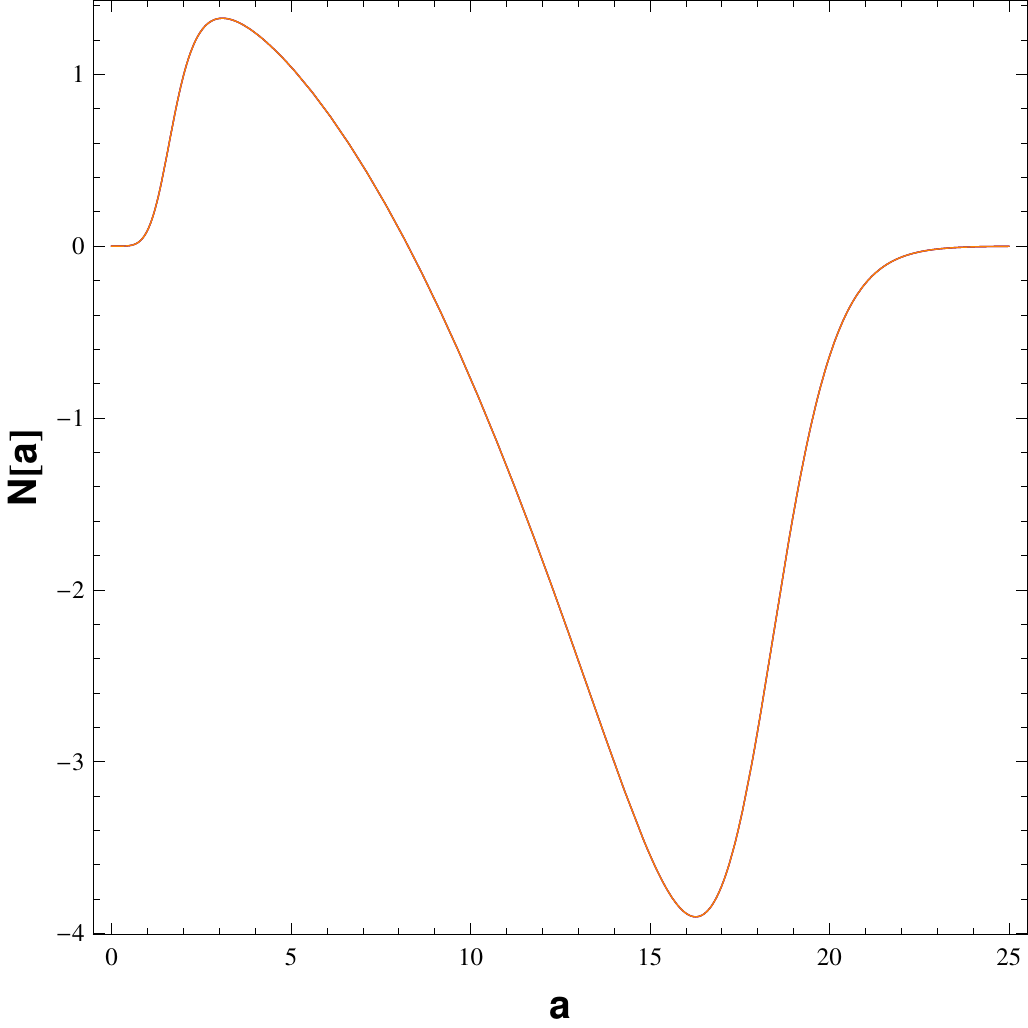}
\caption{Dependence of the non-adiabaticity factor $N$ on the scale factor. The non-adiabaticity is negligible for $a\ll 1$ and for $a\gg 1$. For $a > \sqrt{\frac{3}{2}}\sigma \approx 6$ the dark-energy density becomes negative but this part is suppressed exponentially. The asymptotic value $N=0$ is approached at $a\gtrsim 22$ after $N$ has passed a minimum $N_{min} \approx -4$ at $a \approx 17 $.}
\label{figN}
\end{figure}

Perturbations in an interacting two-component system are necessarily non-adiabatic. The non-adiabatic part of the pressure perturbations is $\hat{p} - \frac{\dot{p}}{\dot{\rho}}\hat{\rho}$. For our special case  the crucial quantity is
\begin{equation}\label{nad}
\hat{p} - \frac{\dot{p}}{\dot{\rho}}\hat{\rho} = \left(c_{s}^{2} + 1\right) \hat{\rho}_{x}^{c} + \frac{\dot{\rho}_{m}\dot{\rho}_{x}}{\dot{\rho}}
\left(\frac{\hat{\rho}^{c}_{m}}{\dot{\rho}_{m}} - \frac{\hat{\rho}_{x}^{c}}{\dot{\rho}_{x}}\right)
= \left(c_{s}^{2} + 1\right) \hat{\rho}_{x}^{c} + \frac{1}{\dot{\rho}}\left[\hat{\rho}^{c}_{m}\dot{\rho}_{x} - \hat{\rho}^{c}_{x}\dot{\rho}_{m}\right]\ .
\end{equation}
Introducing here the balances $\dot{\rho}_{x} = - Q$ and $\dot{\rho}_{m} = - 3H \rho_{m} + Q$
as well as $\delta_{x}^{c} = \epsilon \delta_{x}^{c}$ $\Rightarrow$
$\hat{\rho}^{c}_{x} = \epsilon\frac{\rho_{x}}{\rho_{m}}\hat{\rho}^{c}_{m}$, we find
\begin{equation}\label{nadQ}
\hat{p} - \frac{\dot{p}}{\dot{\rho}}\hat{\rho} = N\hat{\rho}^{c}_{m}\ .
\end{equation}
where
\begin{equation}\label{N}
N \equiv
\left[\frac{Q}{3H\rho_{m}} + \epsilon\frac{\rho_{x}}{\rho_{m}}\left(c_{s}^{2} + \frac{Q}{3H\rho_{m}}\right)\right]\ .
\end{equation}
Even for $\epsilon = 0$, i.e. without fluctuations of the dark-energy component, the interaction term
induces a non-adiabatic contribution to the to total pressure perturbation, given by
(\ref{Q/rhom}) with $g$ from (\ref{dotg+}) and $\frac{\rho_{x}}{\rho_{m}}$ from (\ref{rx/rm}).
The quantity $N$ characterizes the non-adiabatic part of the pressure perturbations. Its dependence on the scale factor is visualized in Fig.~\ref{figN}. For $a > \sqrt{\frac{3}{2}}\sigma \approx 6$ the dark-energy density becomes negative, albeit exponentially suppressed. $N$ approaches its asymptotic value $N=0$ for $a\gtrsim 22$ after passing through a minimum  at $a \approx 17 $. The behavior of $N(a)$
is almost independent of $c_{s}^{2}$ and $\epsilon$.

\section{Summary}
\label{summary}

We have tested a phenomenological model of transient accelerated expansion in which an interaction in the dark sector is constitutive for the cosmological dynamics. The interaction has both to cancel a ``bare" cosmological constant and, at the same time, to generate a phase of accelerated expansion.
While the detailed structure of the interaction was chosen for mathematical convenience, it admits an analytic solution of the background dynamics, we think that it can be used to discuss general features of transient acceleration models. We reconsidered the background dynamics of this model and performed a statistical analysis based on the SNIa data of the Constitution sample. The model predicts a future minimum of the deceleration parameter $q$ which afterwards switches to positive values again (Fig.~\ref{figq}).
The dark-energy density becomes negative for $a\gtrsim 6$ but this contribution is exponentially suppressed. The direction of the energy transfer is from dark energy to dark matter for $a\lesssim 8$.
It is reversed for  $a\gtrsim 8$ but the entire interaction term becomes exponentially suppressed as well.

The perturbation dynamics of the model was investigated both on the Newtonian and on the GR levels. Using a simple parametrization for the perturbed interaction term in the Newtonian setting  and including perturbations of the dark-energy component within a simple ansatz, we carried out a statistical analysis, using the growth-rate data in \cite{gong} and \cite{blake}.
This allowed us to quantify, although with a large dispersion, the role of the perturbed interaction term and the contribution of the dark-energy perturbations. For $a\lesssim 2$ the fractional
matter perturbations deviate stronger from the Einstein-de Sitter behavior than for the $\Lambda$CDM model (Fig.~\ref{figdelta1}). This corresponds to a larger difference for the growth rate $f(z)$ as well. For $a\gtrsim 2$, however, the matter perturbations continue to grow while they approach a constant value for the  $\Lambda$CDM model.
The effective gravitational constant $G_{eff}$ deviates considerably from $G$ for $a>1$ but the ratio $G_{eff}/G$ approaches unity again in the far-future limit (Fig.~\ref{figGeff}).

The relativistic perturbation analysis for the interacting two-component system was performed in terms of gauge-invariant quantities with physical interpretation in the comoving gauge. Perturbations in such type of systems are intrinsically non-adiabatic.  To decouple the relevant perturbation equation we assumed for simplicity that the fractional perturbations of the dark energy are proportional to the fractional matter perturbations. The statistical analysis, based on the data from the 2dFGRS project reveals that the factor of proportionality is very small. In other words, dark-energy perturbations are small on scales that are relevant for structure formation. However, our analysis also shows that the considered data range is not adequately described by a constant factor. For the smallest scales we have much more data than for larger scales. Consequently, the small-scale date have a higher statistical weight than the fewer data on larger scales. On the smallest scales, the dark-energy fluctuations are irrelevant indeed.
On the other hand it is obvious that on larger scales the overall best-fit curve does not provide a good description of the observations. A considerably larger (but still small) value of the mentioned factor
shows a much better performance (Fig.~\ref{Spectrum32}).
This indicates an increasing role of the dark-energy perturbations with increasing scale, a subject that deserves attention in future research.

\acknowledgments{Financial support by CAPES and CNPq is gratefully acknowledged.}


\begin{thebibliography}{99}

\bibitem{riessperl} A. G. Riess et al., Astron. J. 116, 1009
(1998)[astro-ph/9805201 ]; S. J. Perlmutter et al., Astrophys. J.
517, 565(1999); A. G. Riess et al., Astrophys. J. 607, 665(2004);
P. Astier et al., Astron. Astrophys. 447, 31 (2006).

\bibitem{lss}M. Tegmark et al. [SDSS Collaboration], Phys. Rev. D 69, 103501
(2004); K. Abazajian et al. [SDSS Collaboration], Astron. J. 128,
502 (2004); K. Abazajian et al. [SDSS Collaboration], Astron. J.
129, 1755 (2005).

\bibitem{cmb} H. V. Peiris et al., Astrophys. J. Suppl. 148 (2003) 213 [astro-ph/0302225]; C. L.
Bennett et al., Astrophys. J. Suppl. 148  1 (2003); D. N. Spergel
et al., Astrophys. J. Suppl. 148  175 (2003).

\bibitem{isw}
S. Boughn and R. Chrittenden, Nature (London) \textbf{427}, 45
(2004); P. Vielva, E. Mart\'{\i}nez--Gonz\'{a}lez, and M. Tucci,
Mon. Not. R. Astron. Soc. \textbf{365}, 891 (2006).

\bibitem{eisenstein} D.J. Eisenstein {\em et al.},  Ap.J. \textbf{633}, 560 (2005), arXiv:astro-ph/0501171.

\bibitem{weakl}
C.R. Contaldi, H. Hoekstra, and A. Lewis, Phys. Rev. Lett.
\textbf{90}, 221303 (2003).

\bibitem{nieuwenh} T.M. Nieuwenhuizen, P.D. Keefe and V. \v{S}pi\v{c}ka, arXiv:1108.3485.

\bibitem{caldwell} R.R. Caldwell, Phys. Lett. \textbf{B 545}, 23 (2002).

\bibitem{caldwell2} R.R. Caldwell, M. Kamionkowski and N.N.Weinberg, Phys.Rev.Lett. \textbf{91}, 071301 (2003).

    \bibitem{frampton} P.H. Frampton, K.J. Ludwick and R.J. Scherrer, Phys. Rev. D \textbf{84}, 063003 (2011), arXiv:1106.4996.

\bibitem{albrecht} A. Albrecht and C. Skordis, Phys. Rev. Lett. \textbf{84}, 2076 (2000).

\bibitem{barrow} J.D. Barrow, R. Bean and J.Magueijo, Mon. Not. R. Astron. Soc. \textbf{316}, L41 (2000).

\bibitem{bertolami} M.C. Bento, O. Bertolami and N.C. Santos, Phys. Rev. D \textbf{65}, 067301 (2002).

\bibitem{sastaro} A. Shafieloo, V. Sahni and A.A. Starobinsky, Phys. Rev. D \textbf{80}, 101301 (2009), arXiv:0903.5141.

\bibitem{antonio} A.C.C. Guimarães and J.A.S. Lima, Class. Quantum Grav. \textbf{28}, 125026 (2011);
arXiv:1005.2986.

\bibitem{liwuyu} Zhengxiang Li, Puxun Wu and Hongwei Yu, Phys. Lett. \textbf{B695}, 1 (2011); arXiv:1011.1982.

\bibitem{caituo} Rong-Gen Cai and Zhong-Liang Tuo, arXiv:1105.1603.


\bibitem{alcaniz}
F.C. Carvalho, J.S. Alcaniz, J.A.S. Lima and R. Silva, Phys.Rev.Lett. \textbf{97},
081301 (2006), arXiv:astro-ph/0608439.

\bibitem{alcaniztr} F.E M. Costa and J.S. Alcaniz,
arXiv:0908.4251.



\bibitem{transient} J.C. Fabris, B. Fraga, N. Pinto-Neto and W. Zimdahl, JCAP \textbf{1004} (2010) 008,
arXiv:0910.3246.

\bibitem{gong} Yungui Gong, Phys.Rev.D \textbf{78}, 123010 (2008).

\bibitem{blake} Ch. Blake et al., arXiv:1104.2948.

\bibitem{cole} S. Cole et al., Mon. Not. R. Astron. Soc. \textbf{362}, 505 (2005).


\bibitem{sapone} D. Sapone and M. Kunz, Phys.Rev.D \textbf{80}, 083519 (2009).

 \bibitem{Park-Hwang} C.-G. Park, J. Hwang, J. Lee, and H. Noh,
Phys. Rev. Lett. \underline{103}, 151303 (2009).

\bibitem{vernizzi} E. Sefusatti and F. Vernizzi, JCAP \textbf{1103},047 (2011); arXiv:1101.1026.





  \bibitem{saulo} W. Zimdahl, H.A. Borges, S. Carneiro, J.C. Fabris and W.S. Hip\'{o}lito-Ricaldi,
JCAP 1104 (2011) 028.

\bibitem{Hicken} M. Hicken $et$ $al.$,
Astrophys. J. \textbf{700}, 1097 (2009).

\bibitem{diegowang} D. Pav\'{o}n and Bin Wang, Gen. Rel. Grav. {\bf 41}, 1 (2009).

\bibitem{royM} T. Clemson, K. Koyama, Gong-Bo Zhao, R. Maartens and J. Väliviita,
arXiv:1109.6234.

\bibitem{ivan}
I. Dur\'{a}n, D. Pav\'{o}n and W. Zimdahl, JCAP {\bf 1007}(2010) 018.











\bibitem{linderjenkins} E.V. Linder and A. Jenkins, Mon. Not. R. Astron. Soc. \textbf{346}, 573 (2003).

\bibitem{uzan} J.-P. Uzan, Gen.Rel.Grav.\textbf{39}, 307 (2007); arXiv:astro-ph/0605313.

 \bibitem{hutererlinder} D. Huterer and E.V. Linder, Phys.Rev.D \textbf{75}, 023519 (2007); arXiv:astro-ph/0608681.

     \bibitem{diportoamendola} C. Di Porto and L. Amendola, Phys.Rev.D \textbf{77}, 083508 (2008);
    arXiv:0707.2686.


  \bibitem{lahav} A. Kiakotou, {\O}ystein Elgar{\o}y, and O. Lahav, Phys.Rev.D \textbf{77}, 063005 (2008), arXiv:0709.0253

\bibitem{mantz} A. Mantz, S.W. Allen, H. Ebeling and D. Rapetti, Mon. Not. R. Astron. Soc. \textbf{387}, 1179 (2008); arXiv:0709.4294.

\bibitem{nesserisperi} S. Nesseris, L. Perivolaropoulos, Phys.Rev.D \textbf{77}, 023504 (2008); arXiv:0710.1092.



\bibitem{polarski} D. Polarski and R. Gannouji, Phys.Lett.\textbf{B660}, 439 (2008); arXiv:0710.1510



\bibitem{gongishak} Yungui Gong, Mustapha Ishak, Anzhong Wang, Phys.Rev.D \textbf{80}, 023002 (2009); arXiv:0903.0001.

    \bibitem{dosset} J. Dossett, M.Ishak, J. Moldenhauer, Yungui Gong and Anzhong Wang,
         JCAP \textbf{1004} 022 (2010).




     \bibitem{maartensgrowth} G. Caldera-Cabral, R. Maartens and B.M. Schaefer, JCAP \textbf{0907} 027 (2009).





\bibitem{vdf} W.S. Hip\'{o}lito-Ricaldi, H.E.S. Velten and W. Zimdahl,
JCAP 0906 (2009) 016.

\bibitem{bbks} J.M. Bardeen, J.R. Bond, N. Kaiser and A.S. Szalay,
Astrophys. J. {\bf 304}, 15 (1986); J. Martin, A. Riazuelo and M.
Sakellariadou, Phys. Rev. {\bf D61}, 083518 (2000).

\bibitem{sola} J.C. Fabris, I.L. Shapiro and J. Sol\`a, JCAP {\bf
0702}, (2007) 016.
\bibitem{sauloIC} H.A. Borges, S. Carneiro, J.C. Fabris and C.
Pigozzo, Phys. Rev. {\bf D77}, 043513 (2008).

\bibitem{Sandvik} H.B. Sandvik, M. Tegmark, M. Zaldariaga and I. Waga,
Phys. Rev. D 69, 123524 (2004).





\end{thebibliography}
\end{document}